\documentclass[reprint,aps,prb,amsmath,amssymb]{revtex4-2}
\usepackage{mathtools, bm, xcolor, graphicx, hyperref}

\usepackage{multirow}

\usepackage[capitalise]{cleveref}

\usepackage{enumitem}

\begin{document}

\title{$C_3$ symmetry breaking metal-insulator transitions in a flat band in the half-filled Hubbard model on the decorated honeycomb lattice}
\author{H. L. Nourse}
\affiliation{School of Mathematics and Physics, The University of Queensland, Brisbane, Queensland 4072, Australia}
\author{Ross H. McKenzie}
\affiliation{School of Mathematics and Physics, The University of Queensland, Brisbane, Queensland 4072, Australia}
\author{B. J. Powell}
\affiliation{School of Mathematics and Physics, The University of Queensland, Brisbane, Queensland 4072, Australia}

\begin{abstract}
We study the single-orbital Hubbard model on the half-filled decorated honeycomb lattice. In the non-interacting theory at half-filling the Fermi energy lies within a flat band where strong correlations are enhanced. The lattice is highly frustrated. We find a correlation driven first-order metal-insulator transition to two different insulating ground states - a dimer valence bond solid Mott insulator when inter-triangle correlations dominate, and a broken $\mathcal{C}_3$ symmetry antiferromagnet that arises from frustration when intra-triangle correlations dominate. The metal-insulator transitions into these two phases have very different characters. The metal-broken $\mathcal{C}_3$ antiferromagnetic transition is driven by spontaneous $\mathcal{C}_3$ symmetry breaking that lifts the topologically required degeneracy at the Fermi energy and opens an energy gap in the quasiparticle spectrum. The metal-dimer valence bond solid transition breaks no symmetries of the Hamiltonian. It is caused by strong correlations renormalizing the electronic structure into a phase that is adiabatically connected to both the trivial band insulator and the ground state of the spin-1/2 Heisenberg model in the relevant parameter regime. Therefore, neither of these metal-insulator transitions can be understood in either the Brinkmann-Rice or Slater paradigms.
\end{abstract}

\maketitle

\section{Introduction} \label{sec:introduction}

There is increasing interest in flat band systems \cite{Balents2020}. Strongly correlated physics dominates because all states have the same kinetic energy due to the narrow energy range of the flat band. With the recent discovery of strongly correlated insulators \cite{Kim2017,Cao2018,Chen2019,Choi2019,Tang2020,Balents2020} and superconductivity \cite{Cao2018b,Yankowitz2019,Lu2019,Chen2019b,Stepanov2020,Emilio2022,Arora2020,Stepanov2020,Balents2020} in Moir\'e flat bands, an open question is whether the superconductivity is linked to strongly correlated physics and/or the flat bands. On the other hand, there is evidence that some Moir\'e flat bands have superconductivity that is not linked to a correlated insulator \cite{Arora2020,Stepanov2020,Balents2020}. Answering these open questions in strongly correlated physics \cite{Dagotto1994,Imada1998,Orenstein2000,Lee2006,Scalapino2012,Fradkin2015,Keimer2015} with these systems is difficult because the physics in many flat band systems are extremely sensitive to material parameters, making reproducibility and experimental probes difficult. 
%
%It is still unknown to what degree the strongly correlated superconductivity mechanism observed in materials such as the cuprates \cite{Lee2006} or organic superconductors \cite{Powell2011} is linked with the proximate correlated insulator. 
%
Hence, tunable materials that are easy to experimentally probe and that display the above properties are highly sought after.

Many coordination polymers \cite{Batten2013} have elaborate lattices with large geometric frustration, which often results in flat bands similar to the kagome lattice \cite{Murase2017,Murase2017b,Kingsbury2017,Jeon2015,Darago2015,DeGayner2017,Henling2014,Henline2014,Polunin2015,Kalmutzki2018,Jiang2019,Kumar2021}. Coordination polymers are also often strongly correlated, displaying phenomena such as Mott insulators, Kondo physics \cite{Jiang2019,Kumar2021}, and unconventional superconductivity \cite{Zhang2017,Huang2018,Takenaka2021}. Central to our interest is that coordination polymers are highly tunable \cite{Yaghi2016}. Furthermore, coordination polymers often have decorated lattices, and we recently reported that these host a plethora of strongly correlated phenomena as a consequence of the unique lattice and strong electronic correlations \cite{Nourse2021a,Nourse2021b}. 

Of particular interest is the decorated honeycomb lattice, shown in \cref{fig:lattice}a, which hosts many correlated insulators as well as two flat bands. Importantly, the decorated honeycomb lattice has a flat band at half-filling, where a Mott insulator occurs \cite{Nourse2021a}. Additionally, a recent study \cite{Merino2021} reported unconventional strongly correlated superconductivity in the vicinity of this Mott insulator when doping into the flat band within the framework of Anderson's resonating valence bond solid theory \cite{Merino2021}. Hence, decorated lattices in coordination polymers provide a natural path to explore open questions about the connection between strongly correlated insulators, quantum spin liquids, and strongly correlated superconductivity. 

In this study we characterize the ground states of the strongly correlated insulators found in the half-filled Hubbard model on the decorated honeycomb lattice, which is in the vicinity of a flat band. We compare four variational wavefunctions obtained with mean-field rotationally invariant slave bosons (RISB) that incorporate different spatial correlations depending on the cluster choice; a single site cluster, a dimer cluster that exactly captures short range inter-triangle correlations, and a trimer cluster that exactly captures short range intra-triangle correlations. 

\begin{figure}
	\centering
	\includegraphics[width=\columnwidth]{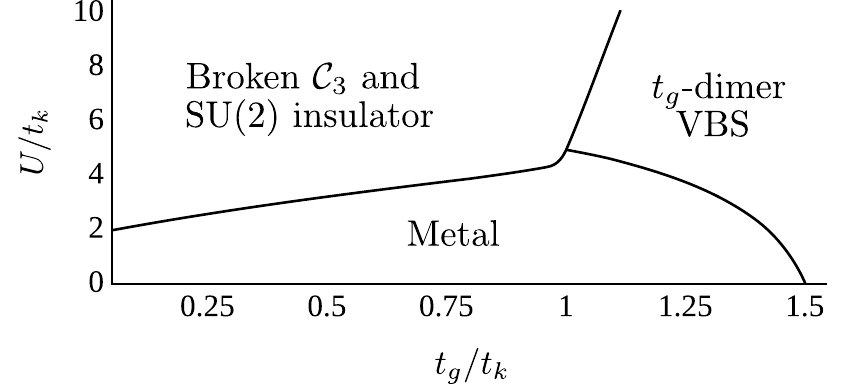}
	\caption
	{\label{fig:phase-diagram}
		Phase diagram of the half-filled Hubbard model on the decorated honeycomb lattice (cf. \cref{fig:lattice}). Strong electronic correlations drive first-order metal-insulator transitions, whose insulating ground state properties depend on the ratio of intra-triangle hopping $t_k$ and inter-triangle hopping $t_g$. For $t_g/t_k \lesssim 1$ in the insulating state the $\mathcal{C}_3$ symmetry of the triangles is broken. Simultaneously each triangle spin polarizes with antiferromagnetic order between triangles (\cref{fig:singlet-schematic}(a)). For $t_g/t_k \gtrsim 1$ the insulating state is instead a valence bond solid (VBS) that forms spin singlets along the $t_g$ bond (\cref{fig:singlet-schematic}(b)). No symmetries are broken in this state.
	}
\end{figure}

The phase diagram of this model is shown in \cref{fig:phase-diagram}. For large enough Coulomb repulsion there are two different insulating phases that occur via uncommon metal-insulator transitions. For strong intra-triangle hopping the frustration of the spins on a triangle causes an insulating state with broken $\mathcal{C}_3$ symmetry (discussed in \cref{sec:broken-c3}). The first-order metal-insulator transition is driven by spontaneous $\mathcal{C}_3$ symmetry breaking, which lifts the topologically required touching of a dispersive band and a flat band at the Fermi energy.
For strong inter-triangle hopping there is a Mott insulator where spin singlets form along the inter-triangle bonds and the ground state is a valence bond solid (VBS) (discussed in \cref{sec:tg-dimer}). In this state no symmetries of the Hamiltonian are broken. The first-order Mott metal-insulator transition occurs via a non-symmetry breaking topological change of the Hamiltonian's electronic structure.
This insulating state is adiabatically connected to both the trivial band insulator and the ground state of the spin-$1/2$ Heisenberg model in the parameter regime where inter-triangle interactions are strong. 
In the large-$U$ limit the ground states compare qualitatively and quantitatively favorably with studies of the Heisenberg model on the decorated honeycomb lattice \cite{Richter2004,Misguich2007,Yang2010,Jahromi2018} (discussed in \cref{sec:real-space-para-heis-literature}).

\section{Model}

\begin{figure}
	\centering
	\includegraphics[width=\columnwidth]{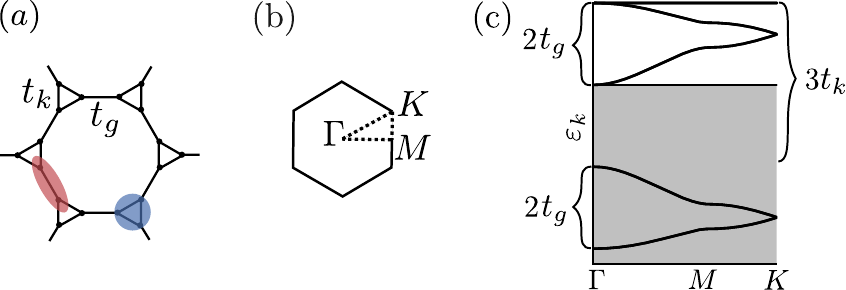}
	\caption
	{\label{fig:lattice}
		(a) The decorated honeycomb lattice. The intra-triangle hopping parameter is $t_k$ and the inter-triangle hopping parameter is $t_g$. The blue oval denotes the three-site trimer cluster and the red oval denotes the two-site dimer cluster. (b) The Brillouin zone, labeling the points of high symmetry. (c) Non-interacting band structure. When there is on average one electron per site (half filling) the Fermi energy lies at the quadratic band touching point at $\Gamma$, with the flat band filled. The gray shaded region denotes the electron filling up to the Fermi energy.
	}
\end{figure}

The single-orbital Hubbard model on the decorated honeycomb lattice is given by
\begin{align} \label{eq:hubbard}
\hat{H} & \equiv -t_g \sum_{\langle i\alpha, j\alpha\rangle,\sigma} \hat{c}_{i\alpha\sigma}^{\dagger} \hat{c}_{j\alpha\sigma}^{} - t_k \sum_{i,\alpha \neq \beta, \sigma} \hat{c}_{i\alpha\sigma}^{\dagger} \hat{c}_{i\beta\sigma}^{} \nonumber \\
& \qquad  + U \sum_{i\alpha} \hat{n}_{i\alpha\uparrow} \hat{n}_{i\alpha\downarrow},
\end{align}
where $\hat{c}_{i\alpha\sigma}^{(\dagger)}$ annihilates (creates) an electron with spin $\sigma$ on site $\alpha$ within triangle $i$, $\hat{n}_{i\alpha\sigma} \equiv \hat{c}_{i\alpha\sigma}^{\dagger} \hat{c}_{i\alpha\sigma}^{}$, $t_g$ ($t_k$) is the hopping parameter between (within) triangles, $U$ is the on-site Coulomb repulsion, and $\langle \cdot \rangle$ denotes only nearest neighbor hopping. We show the lattice in \cref{fig:lattice}a. 

In \cref{fig:lattice}c we show the non-interacting limit for $t_g/t_k \le 3/2$. There are a number of exotic features when the decorated honeycomb lattice is at half filling. The Fermi energy lies at a quadratic band touching point with a flat band at $\vec{k} = \vec{0}$ (the $\Gamma$ point) and the Fermi surface is a single point at $\Gamma$ with two degenerate states \cite{Bergman2008,Jacko2015}. The inclusion of spin orbit coupling opens a gap to a topological quantum spin Hall insulator \cite{Ruegg2010}. At half-filling there is exactly one hole in an infinite peak in the density of states (i.e., one hole shared by the flat-band plus the touching dispersive band at the $\Gamma$ point).

\subsection{Spin-1/2 Heisenberg model}
\label{sec:real-space-para-heis-literature}

In the limit $U \gg t_g,t_k$ the low-energy effective theory of the half-filled single-orbital Hubbard model is the spin-$1/2$ Heisenberg model on the decorated honeycomb lattice, given by
\begin{equation} \label{eq:ham-heis}
	\hat{H}^{\textrm{Heis}} \equiv \frac{J_g}{2} \sum_{\langle i\alpha, j\alpha \rangle} \vec{S}_{i\alpha} \cdot \vec{S}_{j\alpha} + \frac{J_k}{2} \sum_{i,\alpha\neq\beta} \vec{S}_{i\alpha} \cdot \vec{S}_{i\beta},
\end{equation}
where $J_g = 4t_g^2 / U$ is the superexchange along the inter-triangle $t_g$ bonds and $J_k = 4t_k^2 / U$ along the intra-triangle $t_k$ bonds.

Previous exact diagonalization and mean-field studies of the spin-$1/2$ Heisenberg model \cite{Richter2004,Misguich2007,Yang2010} predicts two different VBS states depending on the inter- and intra-spin-exchange coupling of a triangle ($J_g$ and $J_k$ respectively, denoted by $J_e$ and $J_t$ in the cited studies). In the $J_g / J_k \gg 1$ limit (strong coupling between triangles) they predict dimerization along the $J_g$ bond that forms singlets. In the $J_g / J_k \ll 1$ limit (strong coupling within a triangle) they predict a $\sqrt{3} \times \sqrt{3}$ ordered columnar VBS state that forms on a dodecohedron (sixteen-site unit cell) of the decorated honeycomb lattice that breaks the $C_3$ rotational symmetry of a triangle.

An iPEPS study \cite{Jahromi2018} suggests that the $J_g / J_k \ll 1$ limit (strong coupling within a triangle) is instead a VBS with a six-site unit cell, where singlets form along the $J_k$ bonds, but breaks the $C_3$ symmetry of a triangle. They find that the boundary between the $J_g$-dimer VBS and the broken $C_3$-symmetry VBS is $J_g / J_k \sim 0.9$ ($t_g/t_k \sim 0.8$) with three-fold ground state degeneracy.

It is remarkable that our mean-field results with the Hubbard model captures many of the features of the proposed ground states of the Heisenberg model, especially the boundary between the $t_g$-dimer VBS and the broken $\mathcal{C}_3$ symmetry insulator (\cref{fig:phase-diagram}). As we will show in \cref{sec:tg-dimer}, the $t_g$-dimer VBS that we find is likely capturing the correct qualitative ground state electronic configuration, with spin singlets along the $t_g$ bonds and large degeneracy. Conversely, the broken $\mathcal{C}_3$ symmetry state that we discuss in \cref{sec:broken-c3} simultaneously has long range antiferromagnetic order which is not found in the broken $\mathcal{C}_3$ symmetry VBS in \cite{Jahromi2018}. However, our insulating state still has strong short ranged antiferromagnetic correlations. The antiferromagnetic state in our study is likely a consequence of the mean field approximation, which does not treat inter-triangle correlations exactly in the broken $\mathcal{C}_3$ symmetry state.

\section{Method}

We use mean-field RISB \cite{Kotliar1986,Lechermann2007,Lanata2015,Lanata2017} to approximate solutions to \cref{eq:hubbard} using a single-site approximation, two-site dimer clusters (red oval in \cref{fig:lattice}a), and three-site trimer clusters (blue oval in \cref{fig:lattice}a). For all parameters we compare the energy of each cluster solution to obtain the ground state. At the mean-field level RISB is equivalent to the Gutzillwer approximation \cite{Bunemann2007}, which renormalizes a non-interacting wave function by projecting out energetically unfavorable local electronic configurations. Hence, RISB describes the low-energy quasiparticles of a metal and captures metal-insulator transitions.

We implemented mean-field RISB within the \verb+TRIQS+ library \cite{Parcollet2015,Seth2016} at zero temperature. The $k$-integrals were evaluated using the linear tetrahedron method \cite{Blochl1994}, and the ground state of a one bath impurity problem was obtained using exact diagonalization with the Arnoldi method in \verb+ARPACK-NG+ \cite{TRIQS-ARPACK}. We enforced the $\mathcal{C}_3$ symmetry of a triangle and SU(2) symmetry when investigating the Mott insulator with trimer clusters. We relaxed these symmetries when investigating the broken $\mathcal{C}_3$ symmetry solutions. When investigating the dimer solutions we did not enforce any symmetries. Further details of our formalism and implementation are described in \cite{Nourse2021a,Nourse2021b}.

In RISB the physical electron operator is mapped to an enlarged Hilbert space with bosons $\{ \hat{\Phi}_{iAn} \}$ and auxiliary fermions $\{ \hat{f}_{ia} \}$. The bosons keep track of the $2^{M_i}$ quasiparticle electronic configurations $\{ |n_i \rangle \}$ within each cluster $i$ and relates it to the $2^{M_i}$ physical electronic configurations $\{ | A_i \rangle \}$. Here $M_i$ are the number of sites, orbitals, and spins in cluster $i$. We restrict our solutions to the set of bosons where the number of auxiliary fermions in state $| n_i \rangle$, denoted $N_{in}$, is equivalent to the number of physical electrons in state $| A_i \rangle$, denoted $N_{iA}$.

The physical electron in the enlarged Hilbert space is created by
\begin{align}
\underline{\hat{c}}_{i\alpha\sigma}^{\dagger} \equiv \sum_{a\sigma'} \hat{\mathcal{R}}_{ia\alpha}^{\sigma' \sigma} \hat{f}_{ia\sigma'}^{\dagger},
\end{align}
with the unitary operator
\begin{align} \label{eq:R}
\hat{\mathcal{R}}_{ia\alpha}^{\sigma'\sigma} \equiv \sum_{AB} \sum_{nm} \frac{\langle A_i| \hat{c}_{i\alpha\sigma}^{\dagger} | B_i \rangle \langle n_i | \hat{f}_{ia\sigma'}^{\dagger} | m_i \rangle }{\sqrt{N_{iA}(M_i - N_{iB})}} \hat{\Phi}_{iAn}^{\dagger} \hat{\Phi}_{iBm}^{},
\end{align}
where $N_{iA}$ is the number of electrons in state $|A_i \rangle$. The constraints
\begin{align}
\label{eq:constraint1}
& \sum_{An} \hat{\Phi}_{iAn}^{\dagger} \hat{\Phi}_{iAn}^{} = \hat{1}, \\
\label{eq:constraint2}
& \sum_{Anm} \langle m_i | \hat{f}_{ia\sigma}^{\dagger} \hat{f}_{ib\sigma'}^{} |n_i \rangle \hat{\Phi}_{iAn}^{\dagger} \hat{\Phi}_{iAm}^{} = \hat{f}_{ia\sigma}^{\dagger} \hat{f}_{ib\sigma'}^{},
\end{align}
are used to select the physical states out of the enlarged Hilbert space, where $\hat{1}$ is the identity. The first constraint enforces that the physical states are those where there is exactly one boson on every cluster, and the second constraint ensures that the correct boson is attached to the correct auxiliary fermion electronic configuration. Any operator $\hat{X}_i$ on cluster $i$ can be written quadratically in the bosons, given by
\begin{align} \label{eq:local-op-bosons}
\underline{\hat{X}}_i = \sum_{AB} \langle A_i | \hat{X}_i | B_i \rangle \sum_n \hat{\Phi}_{iAn}^{\dagger} \hat{\Phi}_{iAn}^{}.
\end{align}

At the mean-field level we assume all clusters are equivalent and that the bosons condense to a c-number ($\hat{\Phi}_{iAn} \rightarrow \phi_{An}$). Under this approximation the auxiliary fermions are described by an effective non-interacting Hamiltonian of quasiparticles, given by
\begin{align} \label{eq:ham-qp}
	\hat{H}^{\mathrm{qp}} &\equiv -\sum_{ij} \sum_{\alpha\beta,\sigma} \sum_{ab,\sigma'} [\bm{\mathcal{R}}^{}]_{a\alpha}^{\sigma\sigma'} [t_{ij}]_{\alpha\beta}  [\bm{\mathcal{R}}^{\dagger}]_{\beta b}^{\sigma \sigma''} \hat{f}_{ia\sigma'}^{\dagger} \hat{f}_{jb\sigma''}^{} \nonumber \\
	& \phantom{{}\equiv} + \sum_{i,ab,\sigma\sigma'} [\bm{\lambda}^{}]_{ab}^{} \hat{f}_{ia\sigma}^{\dagger} \hat{f}_{ib\sigma'}^{} \nonumber \\
	& = -t_g^* \sum_{\langle ia,ja \rangle,\sigma} \hat{f}_{ia\sigma}^{\dagger} \hat{f}_{ja\sigma}^{} - t_k^* \sum_{i,a\neq b,\sigma} \hat{f}_{ia\sigma}^{\dagger} \hat{f}_{ib\sigma}^{} \nonumber \\
	& \phantom{=} + \ldots,
\end{align}
where $t_g^*$ and $t_k^*$ are the renormalized hopping parameters of \cref{eq:hubbard} and the ellipsis are negligible off-diagonal terms.
Metallic solutions of \cref{eq:ham-qp} at zero temperature describes a Landau Fermi liquid with dressed coherent quasiparticles $\{\hat{f}_{ia\sigma}^{\dagger}\}$ that are renormalized by the local interaction $U$, with the renormalization captured in the mean-field matrices $\bm{\mathcal{R}}$ and $\bm{\lambda}$. \cref{tab:real-space-para} summarizes how these mean-field matrices relate to the renormalized hopping for different cluster choices.

Hence, we we can study the stability of metallic solutions using \cref{eq:ham-qp} from the perspective of band theory with renormalized hopping. Diagonalizing \cref{eq:ham-qp} in reciprocal space gives
\begin{align} \label{eq:qp-bands}
\hat{H}^{\mathrm{qp}} = \sum_{\vec{k}n} \varepsilon_{\vec{k}n}^{\mathrm{qp}} \hat{\psi}_{\vec{k}n}^{\dagger} \hat{\psi}_{\vec{k}n}^{},
\end{align}
where $\varepsilon_{\vec{k}n}^{\mathrm{qp}}$ is the dispersion of the Landau quasiparticles $\{ \hat{\psi}_{\vec{k}n}^{\dagger} \}$ at reciprocal lattice vector $\vec{k}$ and band $n$. On the other hand, from \cref{eq:local-op-bosons}, the local properties on a cluster in real space can be investigated from the condensed bosons.

\subsection{Metal-insulator transitions in RISB}

It is useful to understand how a correlation driven metal-insulator transition is described in the original formulation of Kotliar-Ruckenstein (KR) slave bosons \cite{Kotliar1986}, where a Mott insulator (no symmetry breaking) occurs at half-filling via the Brinkmann-Rice mechanism \cite{Brinkman1970}. As the metal-insulator transition is approached the quasiparticle bands described by \cref{eq:qp-bands} become renormalized and narrow because inter-site hopping is proportional to the quasiparticle weight $Z$. In this case, both $t_k$ and $t_g$ are renormalized by the same amount (\cref{tab:real-space-para}). At the metal-insulator transition the quasiparticle weight vanishes continuously (\cref{fig:real-Z-tg-05}a, dot-dashed green line) and the bandwidth of the quasiparticle bands goes to zero (\cref{fig:real-space-dispersion-tg-05}a). In the Brinkmann-Rice insulator the double occupancy vanishes (\cref{fig:para-d-s}a, dot-dashed green line) with a spin-$1/2$ particle isolated on each site (\cref{fig:para-d-s}b, dot-dashed green line).

Within single-site slave-boson (KR) theory an insulator occurs without breaking a symmetry through the Brinkmann-Rice mechanism, where inter-cluster charge fluctuations vanish, as can be seen from its equivalence to the Gutzwiller approximation \cite{Bunemann2007}. The only other way for a correlated insulator to occur within KR theory is by breaking a symmetry, such as through the Slater mechanism where spontaneous magnetization occurs.

However, because cluster extensions to RISB are able to couple intra-cluster physical electronic configurations with other symmetry compatible quasiparticle configurations (see \cref{eq:R}), the renormalized quasiparticle bands can be shifted and narrowed by differing amounts. Hence, an insulator can occur within RISB without breaking a symmetry that is not through the Brinkmann-Rice mechanism, as has been shown in multi-band extensions to the Gutzwiller approximation \cite{Fabrizio2007}. In \cref{sec:three-site-mott} we discuss a symmetry broken correlated insulator on the half-filled decorated honeycomb lattice, while in \cref{sec:tg-dimer} we discuss a correlated insulator that does not break any symmetries of the Hubbard model and does not occur through the Brinkmann-Rice mechanism.

\section{Broken $\mathcal{C}_3$ symmetry antiferromagnetic insulator}
\label{sec:three-site}

In this section we investigate the half-filled decorated honeycomb lattice in the regime where the trimer cluster (\cref{fig:lattice}a) gives the lowest energy. We find that strong electronic correlations drive a Mott metal-insulator transition, whose low-energy effective theory is the antiferromagnetic spin-$1/2$ Heisenberg model on the decorated honeycomb lattice. The intra-triangle coupling dominates compared to the inter-triangle coupling and the spins within a triangle are frustrated. The system further lowers its energy by simultaneously breaking the $\mathcal{C}_3$ symmetry of a triangle and stabilizing long-range antiferromagnet order. We will describe how RISB captures this state. 

\subsection{Mott insulator (no spontaneous symmetry breaking)}
\label{sec:three-site-mott}

\begin{table} \centering
%\begin{tabular}{@{}cccc@{}}
\begin{tabular}{p{0.20\columnwidth} p{0.23\columnwidth} p{0.23\columnwidth} p{0.23\columnwidth}}
	\multirow{2}{4em}{\centering Hopping parameter} &
	\multicolumn{3}{c}{}    \\
	&
	KR &
	dimer    &
	trimer  \\
	\colrule
	$t_k^*$ \hspace{0.2cm} & $t_k [\bm{Z}]_{11}$ & $t_k [\bm{Z}]_{11}$  & $-[\bm{\lambda}]_{12}$ \\
	$t_g^*$ \hspace{0.2cm} & $t_g [\bm{Z}]_{11}$ & $-[\bm{\lambda}]_{12}$  &  $t_g [\bm{Z}]_{11}$ \\
\end{tabular}
\caption {\label{tab:real-space-para} 
	The components of the auxiliary fermion Hamiltonian that corresponds to the renormalized hopping parameters $t_k^*$ and $t_g^*$ for the three different cluster choices. The quasiparticle weight matrix $\bm{Z} = \bm{\mathcal{R}}^{\dagger} \bm{\mathcal{R}}$ renormalizes inter-cluster hopping, while the correlation potential matrix $\bm{\lambda}$ gives that renormalized intra-cluster hopping. We highlight that $t_g^*$ and $t_k^*$ are approximate descriptions because $\bm{Z}$ is a matrix that may have off-diagonal components. Kotliar-Ruckenstein (KR) refers to a single-site cluster, dimer is a two-site cluster, and trimer is a three-site cluster (cf. \cref{fig:lattice}).
}
\end{table}

We first investigate the correlation driven Mott metal-insulator transition where no symmetries of \cref{eq:hubbard} are broken in order to understand the effects of strong correlations. We enforce the SU(2) and $\mathcal{C}_3$ symmetry of the Hubbard model on the decorated honeycomb lattice. In cluster extensions of mean-field RISB theory the original Hamiltonian \cref{eq:hubbard} is mapped to a non-interacting one with different renormalized inter-triangle $t_g^*$ and intra-triangle $t_k^*$ hopping parameters (\cref{eq:ham-qp}). In the three-site cluster the matrix $\bm{Z} \equiv \bm{\mathcal{R}}^{\dagger} \bm{\mathcal{R}}$ captures the renormalization of $t_g$, while the matrix $\bm{\lambda}$ describes the renormalization of $t_k$ (see \cref{eq:ham-qp,tab:real-space-para}). 

\begin{figure}
	\centering
	\includegraphics[width=\columnwidth]{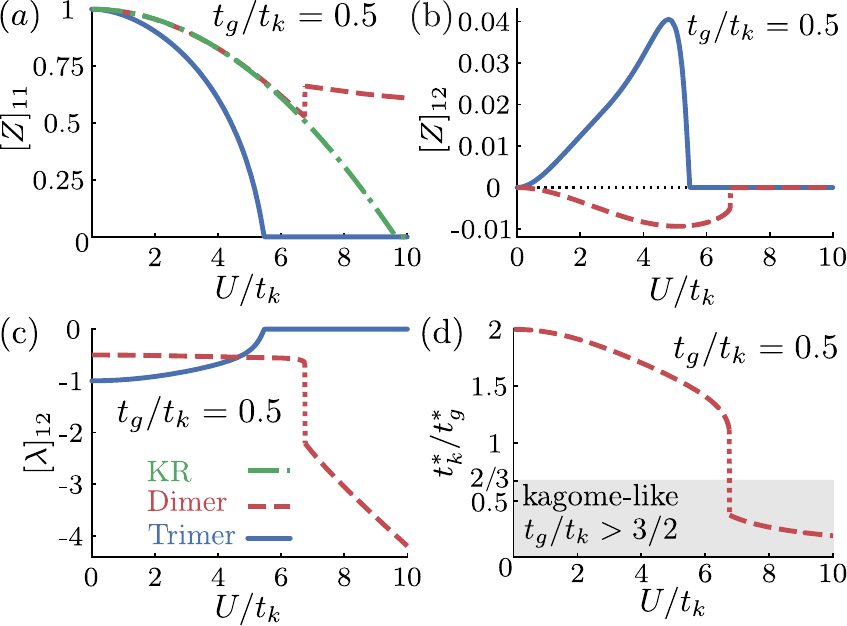}
	\caption
	{\label{fig:real-Z-tg-05} 
		The quasiparticle weight $\bm{Z}$ and correlation potential matrix $\bm{\lambda}$ for $t_g/t_k = 0.5$. We show the KR (single site cluster) solution, the dimer cluster solution, and the trimer cluster solution (cf. \cref{tab:real-space-para}) without allowing any symmetries of the Hamiltonian to break.
		(a) The diagonal components $[\bm{Z}]_{11}$ renormalizes the inter-cluster hopping. In the KR solution the hopping between all sites vanishes at the metal-insulator transition; in the dimer cluster the hopping $t_k^*$ remains finite; and in the trimer cluster hopping between triangles $t_g^*$ vanishes.
		(b) The off-diagonal components $[\bm{Z}]_{12}$ of inter-cluster hopping are small in the correlated metal and vanish at the metal-insulator transition.
		(c) The off-diagonal components of $[\bm{\lambda}]_{12}$ renormalizes the intra-cluster hopping. At the metal-insulator transition in the dimer cluster there is a strong enhancement of $t_g^*$; and in the trimer cluster intra-triangle hopping $t_k^*$ vanishes.
		(d) Renormalized hopping ratio $t_k^* / t_g^*$. The metal-insulator transition occurs in the dimer cluster solution by discontinuously opening a gap and changing the bands to the $t_g / t_k > 3/2$ regime where it is a band insulator.
	}
\end{figure}

\begin{figure}
	\centering
	\includegraphics[width=\columnwidth]{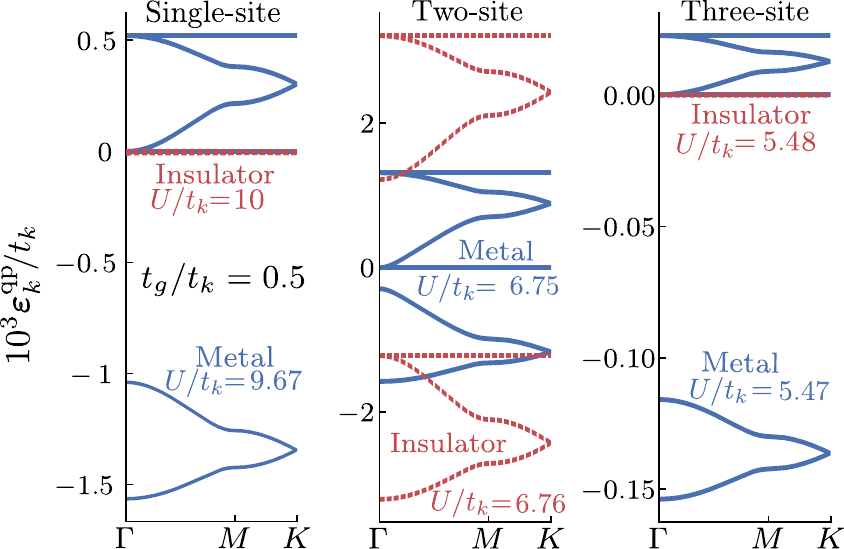}
	\caption
	{\label{fig:real-space-dispersion-tg-05} 
		Renormalized band structure of the KR, dimer cluster, and trimer cluster solutions for $t_g/t_k = 0.5$ without allowing any symmetries of the Hamiltonian to break.
		The Mott insulator on the half-filled decorated honeycomb lattice is captured differently depending on the cluster choice.
		In the (a) KR and (c) trimer cluster solutions the bands narrow and become flat at the metal-insulator transition. In contrast, in the (b) dimer cluster solution the bands become kagome-like ($t_k^*/t_g^* < 3/2$) at the metal-insulator transition and a gap opens.
		The dispersion $\bm{\varepsilon}^{\mathrm{qp}}_k$ denotes the eigenenergies of the quasiparticle Hamiltonian \cref{eq:qp-bands}.
	}
\end{figure}

In \cref{fig:real-Z-tg-05}a (solid blue line) we show the renormalization of the inter-triangle hopping $t_g^* = t_g [\bm{Z}]_{11}$ and in \cref{fig:real-Z-tg-05}c (solid blue line) the renormalized intra-triangle hopping $t_k^* = -[\bm{\lambda}]_{12}$ as electronic correlations are increased. Both $t_g^*$ and $t_k^*$ decrease as the metal-insulator transition is approached until the Fermi surface vanishes. In the insulator $t_g^* = t_k^* = 0$. In \cref{fig:real-space-dispersion-tg-05}c we show the insulator from the perspective of band theory. Correlations narrow the bands. The upper set of bands, which have primarily $E$ molecular orbital character \cite{Nourse2021a,Nourse2021b} narrow more than the lower set of bands, which have primarily $A$ molecular orbital character. This asymmetrical renormalization is captured by small off-diagonal components of the quasiparticle matrix $\bm{Z}$, shown in \cref{fig:real-Z-tg-05}b (solid blue line). Regardless, at the metal-insulator transition the bandwidth of the quasiparticle bands vanishes, indicating electrons have become localized. Hence, the metal-insulator transition occurs by the Brinkmann-Rice mechanism.

\begin{figure}
	\centering
	\includegraphics[width=\columnwidth]{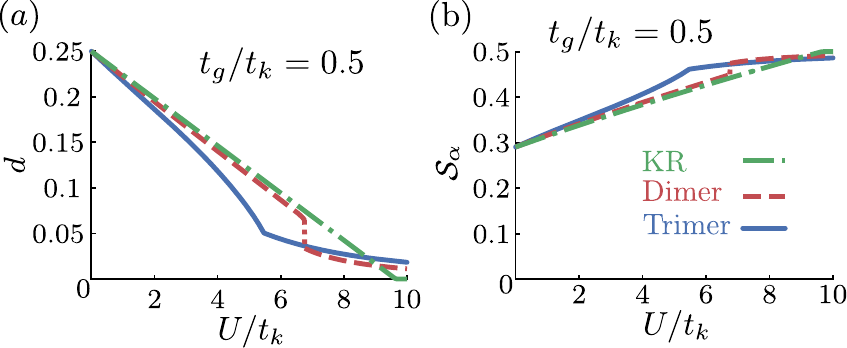}
	\caption
	{\label{fig:para-d-s}
		(a) Double occupancy $d$ and (b) effective spin per site $\mathcal{S}_{\alpha}$ for $t_g/t_k = 0.5$ without allowing any symmetry of the Hamiltonian to break.
		In the insulator there is, on average, one electron localized to a site, acting as a spin-$1/2$ degree of freedom. Only in the two-site and three-site clusters are there inter-site charge and spin fluctuations.
		Double occupancy per site $d$ is given by $d = \sum_{i\alpha} \langle \hat{n}_{i\alpha \uparrow} \hat{n}_{i\alpha \downarrow} \rangle / \mathcal{N}_{\nu}$, where $\mathcal{N}_{\nu}$ is the number of sites on the lattice.
		The effective spin per site is the solution to $\mathcal{S}_{\alpha}(\mathcal{S}_{\alpha} + 1) = \sum_{i\alpha} \langle \vec{S}_{i \alpha} \cdot \vec{S}_{i\alpha} \rangle / \mathcal{N}_{\nu}$ where $\vec{S}_{i\alpha} = \frac{1}{2} \sum_{\sigma \sigma'} \hat{c}_{i\alpha\sigma}^{\dagger} \vec{\tau}_{\sigma \sigma'} \hat{c}_{i\alpha\sigma'}^{}$, and $\vec{\tau}$ is the vector of Pauli matrices.
	}
\end{figure}

In \cref{fig:para-d-s}a (solid blue line) we show the sum of the double occupancy on a site.  In \cref{fig:para-d-s}b (solid blue line) we show the effective spin per site, demonstrating that the localized electrons act as spin-$1/2$ particles. Even though the renormalized hopping parameters vanish in the insulator ($t_g^* = t_k^* = 0$) there are still intra-triangle charge fluctuations. These charge fluctuations are responsible for the dynamical effects in the insulator, such as the superexchange between the spin-$1/2$ particles.

\begin{figure}
	\centering
	\includegraphics[width=\columnwidth]{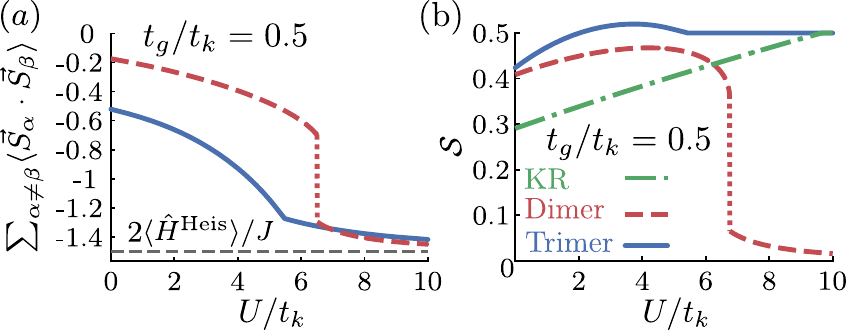}
	\caption
	{\label{fig:para-exchange-s} 
		(a) Sum of the spin-spin correlations within a cluster and (b) total spin $\mathcal{S}$ on a cluster.
		Spin singlets are formed between sites in the insulator. The insulator in the two-site cluster is a dimer lattice with spin-singlets along the inter-triangle $t_g$ bonds. In the three-site cluster spin-singlets form along the intra-triangle $t_k$ bonds. The single-site cluster has no spin exchange between sites because the insulator is adiabatically connected to the atomic limit.
		$\hat{H}^{\textrm{Heis}}$ is the nearest-neighbor spin-$1/2$ Heisenberg model (\cref{eq:ham-heis}) on a $\mathcal{N}_c$ ring with antiferromagnetic exchange.
		The effective spin per cluster is the solution to $\mathcal{S}(\mathcal{S} + 1) = \sum_i \langle \vec{S}_i \cdot \vec{S}_i \rangle \mathcal{N}_c / \mathcal{N}_{\nu}$, where $\vec{S}_i = \sum_{\alpha} \vec{S}_{i\alpha}$, and $\mathcal{N}_c$ is the number of sites within a cluster.
	}
\end{figure}

In \cref{fig:para-exchange-s}a (solid blue line) we show the sum of the intra-triangle spin-exchange between all three sites. In the Mott insulator the total sum of the spin-exchange approaches the limit of an isolated triangle, signifying antiferromagnetic correlation.  Hence, spin-singlets between adjacent sites are favored. However, as shown in \cref{fig:para-exchange-s}b (solid blue line), because of the frustration on a triangle spin-singlets form along the intra-triangle $t_k$ bonds with on average half a spin left over. In \cref{sec:broken-c3} we will show that this frustration causes the $\mathcal{C}_3$ symmetry of a triangle to break and drives a metal-insulator transition. We highlight that in the true ground state there will also be a weaker antiferromagnetic spin-exchange across the $t_g$ bonds of the lattice, but our results only capture this at the mean-field level in the three-site cluster approximation.

The resulting picture is a localized electron on each site  behaving as a spin-$1/2$ particle, with antiferromagnetic spin exchange between sites generated from virtual charge excitations of doubly occupied sites. That is, the Mott insulator is described by the spin-$1/2$ Heisenberg model (\cref{eq:ham-heis}) on the decorated honeycomb lattice, with short range antiferromagnetic correlations within a triangle stabilizing the insulator.

\subsection{Spontaneously broken $\mathcal{C}_3$ symmetry insulator}
\label{sec:broken-c3}

We now relax the constraints to allow the SU(2) and $\mathcal{C}_3$ symmetries of \cref{eq:hubbard} to break. We find that there is a first-order metal-insulator transition where in the insulator each triangle spin polarizes with antiferromagnetic order between triangles and different renormalized hopping $t_k^*$ across the three bonds within a triangle, indicating that the $\mathcal{C}_3$ symmetry of a triangle has broken. The metal-insulator transition is driven by spontaneous $\mathcal{C}_3$ symmetry breaking.

\begin{figure}
	\centering
	\includegraphics[width=\columnwidth]{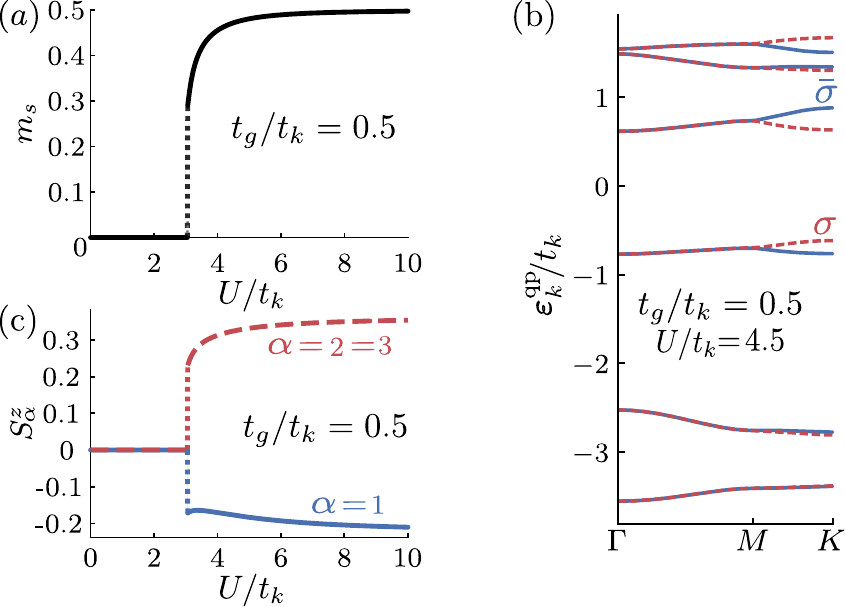}
	\caption
	{\label{fig:afm-sz-bands}
		(a) The staggered magnetization $m_s$ (\cref{eq:ms}). There is a discontinuous metal-insulator transition to an antiferromagnetic state, where the magnetic order is between triangles.
		(b) The quasiparticle bands (\cref{eq:ham-qp}).
		The insulator occurs by spontaneously breaking the $\mathcal{C}_3$ rotational symmetry of a triangle and opening a gap at $\vec{k} = \vec{0}$ (the $\Gamma$ point).
		(c) The average $z$-component of spin for each site in a triangle $S_{\alpha}^z$.
		$S_{\alpha}^z = \sum_i (-1)^{\eta(i)} \sum_{\sigma\sigma'} \langle \hat{n}_{i\alpha\uparrow} - \hat{n}_{i\alpha\downarrow} \rangle / \mathcal{N}_{\nu}$.
		The majority (minority) spin is labeled by $\sigma$ ($\bar{\sigma}$).
	}
\end{figure}

In \cref{fig:afm-sz-bands}a we show the staggered magnetization between triangles, given by
\begin{align} \label{eq:ms}
m_s \equiv \frac{3}{\mathcal{N}_{\nu}} \sum_i \left| (-1)^{\eta(i)} \langle \hat{S}_i^z \rangle \right|,
\end{align}
where $\eta(i) = 0\,(1)$ for the two inequivalent triangles. A finite $m_s$ signifies antiferromagnetic order between triangles due to spin polarization on a triangle and breaks the SU(2) symmetry of the Hamiltonian. There is a discontinuous phase transition to a magnetic state that occurs by increasing $U/t_k$. The state has on average a left over polarized spin-$1/2$ on each triangle ($m_s \rightarrow 1/2$) for large $U$. The antiferromagnetic order lifts the degeneracy of the high symmetry point $K$ (the Dirac point) in the quasiparticle spectrum because inversion about the $t_g$ bond is broken \cite{Nourse2021b}. Correlation driven antiferromagnetism that is responsible for opening an energy gap in the quasiparticle spectrum is known as the Slater mechanism \cite{Slater1951}. However, the Fermi energy is not at the Dirac points and hence the Slater mechanism, driven by antiferromagnetism, is not sufficient to drive the metal-insulator transition.

As was previously discussed in \cref{sec:three-site-mott} for solutions with no broken symmetries, the Mott insulator will favor the formation of spin singlets along the $t_g$ bonds of the lattice. Our calculations does not capture the spin exchange across the $t_g$ bond exactly, the mean-field solution instead describes a broken symmetry state of the singlet, resulting in long-range antiferromagnetic order of the triangles. This suggests that in the true ground state SU(2) may not break and there is instead a singlet along the $t_g$ bond.

Instead, the metal-insulator transition is driven by additionally spontaneously breaking the $\mathcal{C}_3$ symmetry of a triangle. In \cref{fig:afm-sz-bands}c we show the average $z$-component of the spin per site on a triangle. We find that there is a different spin polarization $S_{\alpha}^z$ on each site. Hence, each site is not equivalent and the 120$^{\circ}$ rotational symmetry is broken. Note that this rules out, e.g., 120$^{\circ}$ order of the spins on a triangle. 

The broken $\mathcal{C}_3$ symmetry is accompanied by opening a gap via lifting the two-fold degeneracy at the $\Gamma$ point ($\vec{k}=\vec{0}$) where the flat band touches the dispersive band. In \cref{fig:afm-sz-bands}b we show the quasiparticle spectrum (\cref{eq:qp-bands}) of the magnetic insulator. In the metallic state the Fermi energy is at the quadratic band touching point at $\vec{k} = \vec{0}$. There is one hole per spin flavor shared between the states of the flat band and the state of the dispersive band that touches at the $\Gamma$ point. Because the degeneracy at the $\Gamma$ point is a consequence of the topology of the lattice \cite{Bergman2008,Jacko2015}, the only way to break the degeneracy and open a gap in the quasiparticle spectrum is by breaking symmetries of the Hamiltonian. Spontaneously breaking $\mathcal{C}_3$ symmetry lifts the topological requirement that the flat band touches the dispersive band and allows a gap to open at the Fermi energy. Consequently, in the insulator all quasiparticle bands become isolated from each other, narrow, and highly localized. We are not aware of other examples where a metal-insulator transition occurs via the topologically required degeneracy at the Fermi energy being lifted by strong electronic correlations.

\begin{figure}
	\centering
	\includegraphics[width=\columnwidth]{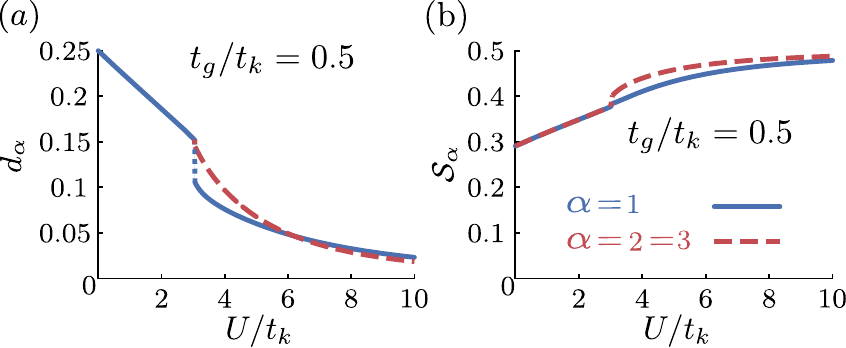}
	\caption
	{\label{fig:af-d-S}
		(a) Double occupancy on each site $d_{\alpha}$ and (b) total spin on each site $\mathcal{S}_{\alpha}$ for the broken $\mathcal{C}_3$ symmetry insulator for $t_g/t_k = 0.5$.
		Increasing correlations causes electrons to localize on each site, with on average one electron per site with small charge fluctuations. For large $U / t_k$ and in the insulator there is on average one electron per site acting as a spin-$1/2$ degree of freedom.
	}
\end{figure}

Similarly to the solutions presented in \cref{sec:three-site-mott}, where no symmetries break, charge fluctuations are heavily suppressed in the broken $\mathcal{C}_3$ symmetry insulator because of strong correlations. In \cref{fig:af-d-S}a we show the double occupancy on a site. Electrons become localized to each site and form an effective spin-$1/2$ degree of freedom (\cref{fig:af-d-S}b) with small charge fluctuations.

\begin{figure}
	\centering
	\includegraphics[width=\columnwidth]{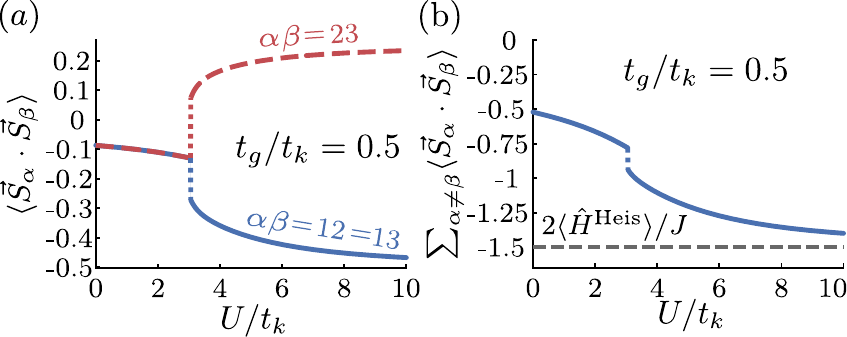}
	\caption
	{\label{fig:afm-correlations}
		(a) Spin-spin correlation between sites within a triangle for the broken $\mathcal{C}_3$ symmetry insulator for $t_g/t_k = 0.5$.
		(b) The sum of the spin correlations between sites within a cluster approaches the correlations in the ground state of the spin-$1/2$ Heisenberg model on an $\mathcal{N}_c$-ring (see \cref{eq:ham-heis}), where $\mathcal{N}_c$ is the number of sites within a cluster.
		The spin operator on site $\alpha$ of cluster $i$ is given by $\vec{S}_{i\alpha} = \frac{1}{2} \sum_{\sigma \sigma'} \hat{c}_{i\alpha\sigma}^{\dagger} \vec{\tau}_{\sigma \sigma'} \hat{c}_{i\alpha\sigma'}^{}$ where $\vec{\tau}$ is the vector of Pauli matrices.
	}
\end{figure}

\begin{figure}
	\centering
	\includegraphics[width=\columnwidth]{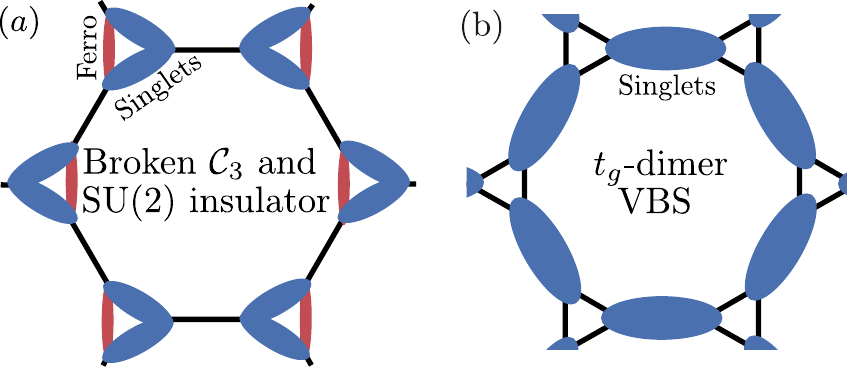}
	\caption
	{\label{fig:singlet-schematic}
		Schematic representation of the singlet formation in the insulators. 
		(a) In the broken $\mathcal{C}_3$ symmetry insulator there are singlets along two of the $t_k$ bonds of a triangle, with the third bond a weaker $S^z = \pm 1$ triplet.
		(b) In the $t_g$-dimer VBS singlets form along the $t_g$ bonds.
		The blue ovals represent singlets and the red ovals represent ferromagnetic correlations. Compare with \cref{fig:afm-correlations}a.
	}
\end{figure}

There are two major differences compared to the Mott insulator with no broken symmetries. First, the spin-$1/2$ degree of freedom on each site polarizes (\cref{fig:afm-sz-bands}b) and the insulator is stabilized by long-range antiferromagnetic correlations, giving antiferromagnetic order between triangles. Second, the equally weighted singlet formation along the $t_k$ bonds within a triangle is broken because of the broken $\mathcal{C}_3$ symmetry. In \cref{fig:afm-correlations}a we show the spin-spin correlations between the sites within a triangle. In the Mott insulator with no broken symmetries frustration makes it difficult to satisfy spin-singlets between neighboring sites within a triangle. In the broken $\mathcal{C}_3$ symmetry insulator the state compromises by polarizing and only anti-aligning two of the sites, with a ferromagnetic spin coupling to the leftover site. We show the short range spin correlations schematically in \cref{fig:singlet-schematic}a. As shown in \cref{fig:afm-correlations}b, the sum of the energy contribution from the spin correlations on a triangle approaches those found in the spin-$1/2$ Heisenberg model on a three-site ring.

\begin{figure}
	\centering
	\includegraphics[width=\columnwidth]{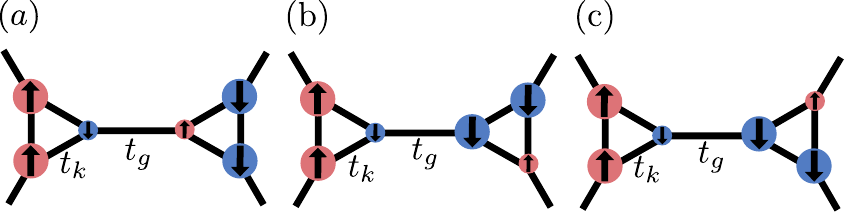}
	\caption
	{\label{fig:afm-schematic-order}
		Schematic representation of the degenerate spin configurations of a site in the broken $\mathcal{C}_3$ insulator.
		There are six other states by permuting the configurations on one of the triangles, and another nine states from flipping the spin on each site. There are in total eighteen degenerate states.
	}
\end{figure}

In \cref{fig:afm-schematic-order} we show a schematic of the magnetic order on the lattice. In our results we showed the representative state shown in \cref{fig:afm-schematic-order}a, but there are other spin configurations that are degenerate. There are a total of eighteen states within a unit cell that have the same energy. However, the macroscopic degeneracy is likely a consequence of the three-site cluster. Correlations are treated exactly along the $t_k$ bonds, while correlations along the $t_g$ bonds are only treated at the mean-field level. Therefore, the superexchange along the $t_g$ bond is not captured faithfully, where it is expected to favor singlet correlations. iPEPS calculations \cite{Jahromi2018}, which we discussed in \cref{sec:real-space-para-heis-literature}, suggests that instead of magnetic order there may be three singlets on a triangle with differing strength. It is likely that our calculations do not capture this state because of the finite cluster size, and hence instead the spins polarize and the SU(2) symmetry of the Hamiltonian breaks. 

%The degeneracy of the ground state may have profound consequences. A Pauling estimate of our antiferromagnetic solutions gives a macroscopic quantum degeneracy of $g \sim 3^{N_s/3}$ (with entropy $S \sim (N_s/3) k_B \ln 3$). The macroscopic degeneracy in our solutions is a consequence of breaking inversion symmetry and being able to freely arrange the polarized spins on each triangle independently of adjacent triangles. In a solution where the SU(2) symmetry does not break but there is instead a singlet along the $t_g$ bond, such a large macroscopic degeneracy implies a resonating valence bond solid (RVB), a quantum spin liquid that may host exotic excitations \cite{Balents2010,Savary2016}. 
%
%However, if the $t_g$ bond is treated exactly then the spin configurations shown in \cref{fig:afm-schematic-order}b,c should have higher energy than the spin configuration shown in \cref{fig:afm-schematic-order}a. The iPEPS results of \cite{Jahromi2018} also breaks inversion symmetry but only has a three fold ground state degeneracy, giving a VBS ground state. If the coupling along the $t_g$ bonds is comparable to the coupling along the $t_k$ bonds, then the singlet correlations may become weaker and the broken $\mathcal{C}_3$ spin-singlet configuration on a triangle may resonate.

\section{$t_g$-dimer valence bond solid Mott insulator}
\label{sec:tg-dimer}

In this section we investigate the effect of strong correlations with a dimer cluster (red oval in \cref{fig:lattice}a) along the $t_g$ bonds, treating the correlations along the $t_g$ bonds exactly and the correlations along the $t_k$ bonds at the mean-field level. We find that there is a true first-order Mott (no broken symmetry of the Hamiltonian) metal-insulator transition to a state that favors spin singlets along the $t_g$ bonds, which we call a $t_g$-dimer VBS. But, that the mechanism is very different from either the Brinkmann-Rice or Slater mechanisms. The mechanism that causes the Mott metal-insulator transition is similar to one that occurs in the Hubbard model on the dimer lattice model \cite{Fabrizio2007}.

In \cref{fig:real-Z-tg-05} (red dashed line) we show the renormalization of the hopping parameters as $U$ is increased for the bare hopping ratio $t_g/t_k = 0.5$. At $U = 0$ the system is metallic with the Fermi energy at the quadratic band touching point at $\vec{k} = \vec{0}$ (the $\Gamma$ point). At a critical $U_c$ there is a discontinuous change of the renormalized hopping parameters $t_k^* / t_g^*$ to the $t_g/t_k > 3 / 2$ regime. At $U = 0$ and $t_g/t_k > 3 / 2$ the system is a band insulator. Electronic correlations drive a metal-insulator transition at $U_c$ by restructuring the electronic structure into the band insulator regime. The resulting insulator is adiabatically connected to the trivial band insulator, and the low-energy excitations remain coherent quasiparticles. Unlike the broken $\mathcal{C}_3$ symmetry insulator, the $t_g$-dimer VBS remains quantum disordered and does not break any symmetries of the Hamiltonian. Instead, the insulator occurs from a symmetry compatible topological change of the Hamiltonian, as is shown in the quasiparticle band structure of \cref{fig:real-space-dispersion-tg-05}b.

In \cref{fig:para-d-s}a (red dashed line) we show the double occupancy per site. Strong electronic correlations suppress charge fluctuations. In the insulator the charge fluctuations become small but do not vanish. Because the $t_g$-dimer VBS does not occur by the Brinkmann-Rice mechanism, there are still inter-dimer charge fluctuations. In \cref{fig:para-d-s}b (red dashed line) we show the effective spin per site and in \cref{fig:para-exchange-s}b (red dashed line) the effective spin per dimer. In the insulator there is, on average, one electron localized to each site acting as a spin-$1/2$ degree of freedom ($\mathcal{S}_{\alpha} \rightarrow 1/2$) with nearest-neighbour antiferromagnetic spin exchange. Hence, a spin singlet forms along the $t_g$ bond ($\mathcal{S} \rightarrow 0$). The resulting picture of the insulator is a VBS with singlets along the $t_g$ bonds, which we schematically show in \cref{fig:singlet-schematic}b.

\section{Boundary between broken $\mathcal{C}_3$ symmetry insulator and $t_g$-dimer VBS}
\label{sec:boundary}

In the phase diagram shown in \cref{fig:phase-diagram} the $t_g$-dimer VBS occurs for $t_g/t_k \gtrsim 1$ for a sufficiently large $U / t_k$. As we have demonstrated, in the half-filled Hubbard model on the decorated honeycomb lattice (\cref{eq:hubbard}) a Mott metal-insulator transition occurs, whose low-energy effective theory is the spin-$1/2$ Heisenberg model (\cref{eq:ham-heis}). An exact solution to the Hubbard model will correctly capture the superexchange between sites. However, within RISB and the clusters we choose the superexchange along the $t_g$ and $t_k$ bonds are not treated on an equal footing.

We can understand the boundary in \cref{fig:phase-diagram} by estimating whether capturing correlations along the $t_g$ or $t_k$ bond is a better representation in the insulating phase by considering the energies of the spin-$1/2$ Heisenberg model on isolated two and three-site rings. The ground state energy for an isolated dimer is $\langle \hat{H}^{\textrm{Heis}} \rangle / J_k = -3/4$ and for an isolated triangle is $\langle \hat{H}^{\textrm{Heis}} \rangle / J_t = -3/4$. Defining $\mathcal{N}_g \equiv \mathcal{N}_{\nu} / 2$ and $\mathcal{N}_k \equiv \mathcal{N}_{\nu} / 3$ as the number of two-site and three-site rings on the decorated honeycomb lattice respectively, an estimation of the ground state energies of the uncoupled clusters of the spin-$1/2$ Heisenberg model is given by
\begin{align}
	E_g & = - \frac{3}{4} \mathcal{N}_g J_g = - \frac{3}{8} \mathcal{N}_{\nu} \frac{4t_g^2}{U}, \nonumber \\
	E_k & = - \frac{3}{4} \mathcal{N}_k J_k = - \frac{1}{4} \mathcal{N}_{\nu} \frac{4t_k^2}{U}.
\end{align}
The energies are equal ($E_g = E_k$) when
\begin{align}
	\frac{3}{8} t_g^2 & = \frac{1}{4} t_k^2 \nonumber \\
	\Rightarrow \frac{t_g}{t_k} & = \sqrt{\frac{2}{3}} \sim 0.816.
\end{align}

The boundary occurs because for $t_g/t_k \gtrsim 0.8$ the system gains more energy by forming singlets along the $t_g$ bond, while for $t_g/t_k \lesssim 0.8$ there is a larger energy gain by forming singlets along the $t_k$ bond. We find the same boundary in our solutions for the Hubbard model where there is no spontaneously broken symmetries. However, as shown in \cref{fig:phase-diagram}, the system can lower its energy further by breaking the $\mathcal{C}_3$ symmetry of a triangle, which extends the trimerized phase to a larger $t_g/t_k$ and the boundary becomes a function of $U/t_k$.

\section{Conclusion} \label{sec:conclusion}

A broken $\mathcal{C}_3$ symmetry insulator and a $t_g$-dimer VBS occurs from strong electronic correlations on the half-filled Hubbard model on the decorated honeycomb lattice at zero temperature.
The metal-to-broken $\mathcal{C}_3$ symmetry insulator occurs via the lifting of the topologically required degeneracy at the $\Gamma$ point and opens an energy gap at the Fermi energy. It is accompanied by long-range antiferromagnetic order between triangles.
The $t_g$-dimer VBS is a Mott insulator where there are no broken symmetries, and is adiabatically connected to both the trivial band insulator and the ground state of the spin-$1/2$ Heisenberg model in the regime where inter-triangle spin-exchange dominates. The ground states found in our electronic mean-field study show remarkable similarities to quantum disordered states in spin models on the decorated honeycomb lattice \cite{Richter2004,Misguich2007,Yang2010,Jahromi2018}. 

Importantly, the insulating states we find occur at half-filling of a flat band, where strong correlations are typically enhanced. With the recent prediction of unconventional superconductivity on the decorated honeycomb lattice near half-filling \cite{Merino2021}, there is an open question about the connection between the strongly correlated insulator found in our study and unconventional superconductivity. A useful direction to explore this connection may be in coordination complexes and polymers where the decorated honeycomb lattice is often found \cite{Murase2017,Murase2017b,Kingsbury2017,Jeon2015,Darago2015,DeGayner2017,Henling2014,Henline2014,Polunin2015,Kalmutzki2018,Jiang2019,Kumar2021}.

\begin{acknowledgments}
This work was supported by the Australian Research Council through Grant No. DP181006201.
\end{acknowledgments}

%\appendix

%\section{appendix}
%appendix

%\bibliography{bibliography}

\begin{thebibliography}{59}%
	\makeatletter
	\providecommand \@ifxundefined [1]{%
		\@ifx{#1\undefined}
	}%
	\providecommand \@ifnum [1]{%
		\ifnum #1\expandafter \@firstoftwo
		\else \expandafter \@secondoftwo
		\fi
	}%
	\providecommand \@ifx [1]{%
		\ifx #1\expandafter \@firstoftwo
		\else \expandafter \@secondoftwo
		\fi
	}%
	\providecommand \natexlab [1]{#1}%
	\providecommand \enquote  [1]{``#1''}%
	\providecommand \bibnamefont  [1]{#1}%
	\providecommand \bibfnamefont [1]{#1}%
	\providecommand \citenamefont [1]{#1}%
	\providecommand \href@noop [0]{\@secondoftwo}%
	\providecommand \href [0]{\begingroup \@sanitize@url \@href}%
	\providecommand \@href[1]{\@@startlink{#1}\@@href}%
	\providecommand \@@href[1]{\endgroup#1\@@endlink}%
	\providecommand \@sanitize@url [0]{\catcode `\\12\catcode `\$12\catcode
		`\&12\catcode `\#12\catcode `\^12\catcode `\_12\catcode `\%12\relax}%
	\providecommand \@@startlink[1]{}%
	\providecommand \@@endlink[0]{}%
	\providecommand \url  [0]{\begingroup\@sanitize@url \@url }%
	\providecommand \@url [1]{\endgroup\@href {#1}{\urlprefix }}%
	\providecommand \urlprefix  [0]{URL }%
	\providecommand \Eprint [0]{\href }%
	\providecommand \doibase [0]{https://doi.org/}%
	\providecommand \selectlanguage [0]{\@gobble}%
	\providecommand \bibinfo  [0]{\@secondoftwo}%
	\providecommand \bibfield  [0]{\@secondoftwo}%
	\providecommand \translation [1]{[#1]}%
	\providecommand \BibitemOpen [0]{}%
	\providecommand \bibitemStop [0]{}%
	\providecommand \bibitemNoStop [0]{.\EOS\space}%
	\providecommand \EOS [0]{\spacefactor3000\relax}%
	\providecommand \BibitemShut  [1]{\csname bibitem#1\endcsname}%
	\let\auto@bib@innerbib\@empty
	%</preamble>
	\bibitem [{\citenamefont {Balents}\ \emph {et~al.}(2020)\citenamefont
		{Balents}, \citenamefont {Dean}, \citenamefont {Efetov},\ and\ \citenamefont
		{Young}}]{Balents2020}%
	\BibitemOpen
	\bibfield  {author} {\bibinfo {author} {\bibfnamefont {L.}~\bibnamefont
			{Balents}}, \bibinfo {author} {\bibfnamefont {C.~R.}\ \bibnamefont {Dean}},
		\bibinfo {author} {\bibfnamefont {D.~K.}\ \bibnamefont {Efetov}},\ and\
		\bibinfo {author} {\bibfnamefont {A.~F.}\ \bibnamefont {Young}},\ }\href
	{https://doi.org/10.1038/s41567-020-0906-9} {\bibfield  {journal} {\bibinfo
			{journal} {Nat. Phys.}\ }\textbf {\bibinfo {volume} {16}},\ \bibinfo {pages}
		{725} (\bibinfo {year} {2020})}\BibitemShut {NoStop}%
	\bibitem [{\citenamefont {Kim}\ \emph {et~al.}(2017)\citenamefont {Kim},
		\citenamefont {DaSilva}, \citenamefont {Huang}, \citenamefont {Fallahazad},
		\citenamefont {Larentis}, \citenamefont {Taniguchi}, \citenamefont
		{Watanabe}, \citenamefont {LeRoy}, \citenamefont {MacDonald},\ and\
		\citenamefont {Tutuc}}]{Kim2017}%
	\BibitemOpen
	\bibfield  {author} {\bibinfo {author} {\bibfnamefont {K.}~\bibnamefont
			{Kim}}, \bibinfo {author} {\bibfnamefont {A.}~\bibnamefont {DaSilva}},
		\bibinfo {author} {\bibfnamefont {S.}~\bibnamefont {Huang}}, \bibinfo
		{author} {\bibfnamefont {B.}~\bibnamefont {Fallahazad}}, \bibinfo {author}
		{\bibfnamefont {S.}~\bibnamefont {Larentis}}, \bibinfo {author}
		{\bibfnamefont {T.}~\bibnamefont {Taniguchi}}, \bibinfo {author}
		{\bibfnamefont {K.}~\bibnamefont {Watanabe}}, \bibinfo {author}
		{\bibfnamefont {B.~J.}\ \bibnamefont {LeRoy}}, \bibinfo {author}
		{\bibfnamefont {A.~H.}\ \bibnamefont {MacDonald}},\ and\ \bibinfo {author}
		{\bibfnamefont {E.}~\bibnamefont {Tutuc}},\ }\href
	{https://doi.org/10.1073/pnas.1620140114} {\bibfield  {journal} {\bibinfo
			{journal} {Proc. Natl. Acad. Sci. U.S.A.}\ }\textbf {\bibinfo {volume}
			{114}},\ \bibinfo {pages} {3364} (\bibinfo {year} {2017})}\BibitemShut
	{NoStop}%
	\bibitem [{\citenamefont {Cao}\ \emph {et~al.}(2018{\natexlab{a}})\citenamefont
		{Cao}, \citenamefont {Fatemi}, \citenamefont {Demir}, \citenamefont {Fang},
		\citenamefont {Tomarken}, \citenamefont {Luo}, \citenamefont
		{Sanchez-Yamagishi}, \citenamefont {Watanabe}, \citenamefont {Taniguchi},
		\citenamefont {Kaxiras}, \citenamefont {Ashoori},\ and\ \citenamefont
		{Jarillo-Herrero}}]{Cao2018}%
	\BibitemOpen
	\bibfield  {author} {\bibinfo {author} {\bibfnamefont {Y.}~\bibnamefont
			{Cao}}, \bibinfo {author} {\bibfnamefont {V.}~\bibnamefont {Fatemi}},
		\bibinfo {author} {\bibfnamefont {A.}~\bibnamefont {Demir}}, \bibinfo
		{author} {\bibfnamefont {S.}~\bibnamefont {Fang}}, \bibinfo {author}
		{\bibfnamefont {S.~L.}\ \bibnamefont {Tomarken}}, \bibinfo {author}
		{\bibfnamefont {J.~Y.}\ \bibnamefont {Luo}}, \bibinfo {author} {\bibfnamefont
			{J.~D.}\ \bibnamefont {Sanchez-Yamagishi}}, \bibinfo {author} {\bibfnamefont
			{K.}~\bibnamefont {Watanabe}}, \bibinfo {author} {\bibfnamefont
			{T.}~\bibnamefont {Taniguchi}}, \bibinfo {author} {\bibfnamefont
			{E.}~\bibnamefont {Kaxiras}}, \bibinfo {author} {\bibfnamefont {R.~C.}\
			\bibnamefont {Ashoori}},\ and\ \bibinfo {author} {\bibfnamefont
			{P.}~\bibnamefont {Jarillo-Herrero}},\ }\href
	{https://doi.org/10.1038/nature26154} {\bibfield  {journal} {\bibinfo
			{journal} {Nature}\ }\textbf {\bibinfo {volume} {556}},\ \bibinfo {pages}
		{80} (\bibinfo {year} {2018}{\natexlab{a}})}\BibitemShut {NoStop}%
	\bibitem [{\citenamefont {Chen}\ \emph
		{et~al.}(2019{\natexlab{a}})\citenamefont {Chen}, \citenamefont {Jiang},
		\citenamefont {Wu}, \citenamefont {Lyu}, \citenamefont {Li}, \citenamefont
		{Chittari}, \citenamefont {Watanabe}, \citenamefont {Taniguchi},
		\citenamefont {Shi}, \citenamefont {Jung}, \citenamefont {Zhang},\ and\
		\citenamefont {Wang}}]{Chen2019}%
	\BibitemOpen
	\bibfield  {author} {\bibinfo {author} {\bibfnamefont {G.}~\bibnamefont
			{Chen}}, \bibinfo {author} {\bibfnamefont {L.}~\bibnamefont {Jiang}},
		\bibinfo {author} {\bibfnamefont {S.}~\bibnamefont {Wu}}, \bibinfo {author}
		{\bibfnamefont {B.}~\bibnamefont {Lyu}}, \bibinfo {author} {\bibfnamefont
			{H.}~\bibnamefont {Li}}, \bibinfo {author} {\bibfnamefont {B.~L.}\
			\bibnamefont {Chittari}}, \bibinfo {author} {\bibfnamefont {K.}~\bibnamefont
			{Watanabe}}, \bibinfo {author} {\bibfnamefont {T.}~\bibnamefont {Taniguchi}},
		\bibinfo {author} {\bibfnamefont {Z.}~\bibnamefont {Shi}}, \bibinfo {author}
		{\bibfnamefont {J.}~\bibnamefont {Jung}}, \bibinfo {author} {\bibfnamefont
			{Y.}~\bibnamefont {Zhang}},\ and\ \bibinfo {author} {\bibfnamefont
			{F.}~\bibnamefont {Wang}},\ }\href
	{https://doi.org/10.1038/s41567-018-0387-2} {\bibfield  {journal} {\bibinfo
			{journal} {Nat. Phys.}\ }\textbf {\bibinfo {volume} {15}},\ \bibinfo {pages}
		{237} (\bibinfo {year} {2019}{\natexlab{a}})}\BibitemShut {NoStop}%
	\bibitem [{\citenamefont {Choi}\ \emph {et~al.}(2019)\citenamefont {Choi},
		\citenamefont {Kemmer}, \citenamefont {Peng}, \citenamefont {Thomson},
		\citenamefont {Arora}, \citenamefont {Polski}, \citenamefont {Zhang},
		\citenamefont {Ren}, \citenamefont {Alicea}, \citenamefont {Refael},
		\citenamefont {von Oppen}, \citenamefont {Watanabe}, \citenamefont
		{Taniguchi},\ and\ \citenamefont {Nadj-Perge}}]{Choi2019}%
	\BibitemOpen
	\bibfield  {author} {\bibinfo {author} {\bibfnamefont {Y.}~\bibnamefont
			{Choi}}, \bibinfo {author} {\bibfnamefont {J.}~\bibnamefont {Kemmer}},
		\bibinfo {author} {\bibfnamefont {Y.}~\bibnamefont {Peng}}, \bibinfo {author}
		{\bibfnamefont {A.}~\bibnamefont {Thomson}}, \bibinfo {author} {\bibfnamefont
			{H.}~\bibnamefont {Arora}}, \bibinfo {author} {\bibfnamefont
			{R.}~\bibnamefont {Polski}}, \bibinfo {author} {\bibfnamefont
			{Y.}~\bibnamefont {Zhang}}, \bibinfo {author} {\bibfnamefont
			{H.}~\bibnamefont {Ren}}, \bibinfo {author} {\bibfnamefont {J.}~\bibnamefont
			{Alicea}}, \bibinfo {author} {\bibfnamefont {G.}~\bibnamefont {Refael}},
		\bibinfo {author} {\bibfnamefont {F.}~\bibnamefont {von Oppen}}, \bibinfo
		{author} {\bibfnamefont {K.}~\bibnamefont {Watanabe}}, \bibinfo {author}
		{\bibfnamefont {T.}~\bibnamefont {Taniguchi}},\ and\ \bibinfo {author}
		{\bibfnamefont {S.}~\bibnamefont {Nadj-Perge}},\ }\href
	{https://doi.org/10.1038/s41567-019-0606-5} {\bibfield  {journal} {\bibinfo
			{journal} {Nat. Phys.}\ }\textbf {\bibinfo {volume} {15}},\ \bibinfo {pages}
		{1174} (\bibinfo {year} {2019})}\BibitemShut {NoStop}%
	\bibitem [{\citenamefont {Tang}\ \emph {et~al.}(2020)\citenamefont {Tang},
		\citenamefont {Li}, \citenamefont {Li}, \citenamefont {Xu}, \citenamefont
		{Liu}, \citenamefont {Barmak}, \citenamefont {Watanabe}, \citenamefont
		{Taniguchi}, \citenamefont {MacDonald}, \citenamefont {Shan},\ and\
		\citenamefont {Mak}}]{Tang2020}%
	\BibitemOpen
	\bibfield  {author} {\bibinfo {author} {\bibfnamefont {Y.}~\bibnamefont
			{Tang}}, \bibinfo {author} {\bibfnamefont {L.}~\bibnamefont {Li}}, \bibinfo
		{author} {\bibfnamefont {T.}~\bibnamefont {Li}}, \bibinfo {author}
		{\bibfnamefont {Y.}~\bibnamefont {Xu}}, \bibinfo {author} {\bibfnamefont
			{S.}~\bibnamefont {Liu}}, \bibinfo {author} {\bibfnamefont {K.}~\bibnamefont
			{Barmak}}, \bibinfo {author} {\bibfnamefont {K.}~\bibnamefont {Watanabe}},
		\bibinfo {author} {\bibfnamefont {T.}~\bibnamefont {Taniguchi}}, \bibinfo
		{author} {\bibfnamefont {A.~H.}\ \bibnamefont {MacDonald}}, \bibinfo {author}
		{\bibfnamefont {J.}~\bibnamefont {Shan}},\ and\ \bibinfo {author}
		{\bibfnamefont {K.~F.}\ \bibnamefont {Mak}},\ }\href
	{https://doi.org/10.1038/s41586-020-2085-3} {\bibfield  {journal} {\bibinfo
			{journal} {Nature}\ }\textbf {\bibinfo {volume} {579}},\ \bibinfo {pages}
		{353} (\bibinfo {year} {2020})}\BibitemShut {NoStop}%
	\bibitem [{\citenamefont {Cao}\ \emph {et~al.}(2018{\natexlab{b}})\citenamefont
		{Cao}, \citenamefont {Fatemi}, \citenamefont {Fang}, \citenamefont
		{Watanabe}, \citenamefont {Taniguchi}, \citenamefont {Kaxiras},\ and\
		\citenamefont {Jarillo-Herrero}}]{Cao2018b}%
	\BibitemOpen
	\bibfield  {author} {\bibinfo {author} {\bibfnamefont {Y.}~\bibnamefont
			{Cao}}, \bibinfo {author} {\bibfnamefont {V.}~\bibnamefont {Fatemi}},
		\bibinfo {author} {\bibfnamefont {S.}~\bibnamefont {Fang}}, \bibinfo {author}
		{\bibfnamefont {K.}~\bibnamefont {Watanabe}}, \bibinfo {author}
		{\bibfnamefont {T.}~\bibnamefont {Taniguchi}}, \bibinfo {author}
		{\bibfnamefont {E.}~\bibnamefont {Kaxiras}},\ and\ \bibinfo {author}
		{\bibfnamefont {P.}~\bibnamefont {Jarillo-Herrero}},\ }\href
	{https://doi.org/10.1038/nature26160} {\bibfield  {journal} {\bibinfo
			{journal} {Nature}\ }\textbf {\bibinfo {volume} {556}},\ \bibinfo {pages}
		{43} (\bibinfo {year} {2018}{\natexlab{b}})}\BibitemShut {NoStop}%
	\bibitem [{\citenamefont {Yankowitz}\ \emph {et~al.}(2019)\citenamefont
		{Yankowitz}, \citenamefont {Chen}, \citenamefont {Polshyn}, \citenamefont
		{Zhang}, \citenamefont {Watanabe}, \citenamefont {Taniguchi}, \citenamefont
		{Graf}, \citenamefont {Young},\ and\ \citenamefont {Dean}}]{Yankowitz2019}%
	\BibitemOpen
	\bibfield  {author} {\bibinfo {author} {\bibfnamefont {M.}~\bibnamefont
			{Yankowitz}}, \bibinfo {author} {\bibfnamefont {S.}~\bibnamefont {Chen}},
		\bibinfo {author} {\bibfnamefont {H.}~\bibnamefont {Polshyn}}, \bibinfo
		{author} {\bibfnamefont {Y.}~\bibnamefont {Zhang}}, \bibinfo {author}
		{\bibfnamefont {K.}~\bibnamefont {Watanabe}}, \bibinfo {author}
		{\bibfnamefont {T.}~\bibnamefont {Taniguchi}}, \bibinfo {author}
		{\bibfnamefont {D.}~\bibnamefont {Graf}}, \bibinfo {author} {\bibfnamefont
			{A.~F.}\ \bibnamefont {Young}},\ and\ \bibinfo {author} {\bibfnamefont
			{C.~R.}\ \bibnamefont {Dean}},\ }\href
	{https://doi.org/10.1126/science.aav1910} {\bibfield  {journal} {\bibinfo
			{journal} {Science}\ }\textbf {\bibinfo {volume} {363}},\ \bibinfo {pages}
		{1059} (\bibinfo {year} {2019})}\BibitemShut {NoStop}%
	\bibitem [{\citenamefont {Lu}\ \emph {et~al.}(2019)\citenamefont {Lu},
		\citenamefont {Stepanov}, \citenamefont {Yang}, \citenamefont {Xie},
		\citenamefont {Aamir}, \citenamefont {Das}, \citenamefont {Urgell},
		\citenamefont {Watanabe}, \citenamefont {Taniguchi}, \citenamefont {Zhang},
		\citenamefont {Bachtold}, \citenamefont {MacDonald},\ and\ \citenamefont
		{Efetov}}]{Lu2019}%
	\BibitemOpen
	\bibfield  {author} {\bibinfo {author} {\bibfnamefont {X.}~\bibnamefont
			{Lu}}, \bibinfo {author} {\bibfnamefont {P.}~\bibnamefont {Stepanov}},
		\bibinfo {author} {\bibfnamefont {W.}~\bibnamefont {Yang}}, \bibinfo {author}
		{\bibfnamefont {M.}~\bibnamefont {Xie}}, \bibinfo {author} {\bibfnamefont
			{M.~A.}\ \bibnamefont {Aamir}}, \bibinfo {author} {\bibfnamefont
			{I.}~\bibnamefont {Das}}, \bibinfo {author} {\bibfnamefont {C.}~\bibnamefont
			{Urgell}}, \bibinfo {author} {\bibfnamefont {K.}~\bibnamefont {Watanabe}},
		\bibinfo {author} {\bibfnamefont {T.}~\bibnamefont {Taniguchi}}, \bibinfo
		{author} {\bibfnamefont {G.}~\bibnamefont {Zhang}}, \bibinfo {author}
		{\bibfnamefont {A.}~\bibnamefont {Bachtold}}, \bibinfo {author}
		{\bibfnamefont {A.~H.}\ \bibnamefont {MacDonald}},\ and\ \bibinfo {author}
		{\bibfnamefont {D.~K.}\ \bibnamefont {Efetov}},\ }\href
	{https://doi.org/10.1038/s41586-019-1695-0} {\bibfield  {journal} {\bibinfo
			{journal} {Nature}\ }\textbf {\bibinfo {volume} {574}},\ \bibinfo {pages}
		{653} (\bibinfo {year} {2019})}\BibitemShut {NoStop}%
	\bibitem [{\citenamefont {Chen}\ \emph
		{et~al.}(2019{\natexlab{b}})\citenamefont {Chen}, \citenamefont {Sharpe},
		\citenamefont {Gallagher}, \citenamefont {Rosen}, \citenamefont {Fox},
		\citenamefont {Jiang}, \citenamefont {Lyu}, \citenamefont {Li}, \citenamefont
		{Watanabe}, \citenamefont {Taniguchi}, \citenamefont {Jung}, \citenamefont
		{Shi}, \citenamefont {Goldhaber-Gordon}, \citenamefont {Zhang},\ and\
		\citenamefont {Wang}}]{Chen2019b}%
	\BibitemOpen
	\bibfield  {author} {\bibinfo {author} {\bibfnamefont {G.}~\bibnamefont
			{Chen}}, \bibinfo {author} {\bibfnamefont {A.~L.}\ \bibnamefont {Sharpe}},
		\bibinfo {author} {\bibfnamefont {P.}~\bibnamefont {Gallagher}}, \bibinfo
		{author} {\bibfnamefont {I.~T.}\ \bibnamefont {Rosen}}, \bibinfo {author}
		{\bibfnamefont {E.~J.}\ \bibnamefont {Fox}}, \bibinfo {author} {\bibfnamefont
			{L.}~\bibnamefont {Jiang}}, \bibinfo {author} {\bibfnamefont
			{B.}~\bibnamefont {Lyu}}, \bibinfo {author} {\bibfnamefont {H.}~\bibnamefont
			{Li}}, \bibinfo {author} {\bibfnamefont {K.}~\bibnamefont {Watanabe}},
		\bibinfo {author} {\bibfnamefont {T.}~\bibnamefont {Taniguchi}}, \bibinfo
		{author} {\bibfnamefont {J.}~\bibnamefont {Jung}}, \bibinfo {author}
		{\bibfnamefont {Z.}~\bibnamefont {Shi}}, \bibinfo {author} {\bibfnamefont
			{D.}~\bibnamefont {Goldhaber-Gordon}}, \bibinfo {author} {\bibfnamefont
			{Y.}~\bibnamefont {Zhang}},\ and\ \bibinfo {author} {\bibfnamefont
			{F.}~\bibnamefont {Wang}},\ }\href
	{https://doi.org/10.1038/s41586-019-1393-y} {\bibfield  {journal} {\bibinfo
			{journal} {Nature}\ }\textbf {\bibinfo {volume} {572}},\ \bibinfo {pages}
		{215} (\bibinfo {year} {2019}{\natexlab{b}})}\BibitemShut {NoStop}%
	\bibitem [{\citenamefont {Stepanov}\ \emph {et~al.}(2020)\citenamefont
		{Stepanov}, \citenamefont {Das}, \citenamefont {Lu}, \citenamefont
		{Fahimniya}, \citenamefont {Watanabe}, \citenamefont {Taniguchi},
		\citenamefont {Koppens}, \citenamefont {Lischner}, \citenamefont {Levitov},\
		and\ \citenamefont {Efetov}}]{Stepanov2020}%
	\BibitemOpen
	\bibfield  {author} {\bibinfo {author} {\bibfnamefont {P.}~\bibnamefont
			{Stepanov}}, \bibinfo {author} {\bibfnamefont {I.}~\bibnamefont {Das}},
		\bibinfo {author} {\bibfnamefont {X.}~\bibnamefont {Lu}}, \bibinfo {author}
		{\bibfnamefont {A.}~\bibnamefont {Fahimniya}}, \bibinfo {author}
		{\bibfnamefont {K.}~\bibnamefont {Watanabe}}, \bibinfo {author}
		{\bibfnamefont {T.}~\bibnamefont {Taniguchi}}, \bibinfo {author}
		{\bibfnamefont {F.~H.~L.}\ \bibnamefont {Koppens}}, \bibinfo {author}
		{\bibfnamefont {J.}~\bibnamefont {Lischner}}, \bibinfo {author}
		{\bibfnamefont {L.}~\bibnamefont {Levitov}},\ and\ \bibinfo {author}
		{\bibfnamefont {D.~K.}\ \bibnamefont {Efetov}},\ }\href
	{https://doi.org/10.1038/s41586-020-2459-6} {\bibfield  {journal} {\bibinfo
			{journal} {Nature}\ }\textbf {\bibinfo {volume} {583}},\ \bibinfo {pages}
		{375} (\bibinfo {year} {2020})}\BibitemShut {NoStop}%
	\bibitem [{\citenamefont {Emilio}\ \emph {et~al.}(2022)\citenamefont {Emilio},
		\citenamefont {Qiyue}, \citenamefont {Ryan}, \citenamefont {Shi},
		\citenamefont {Haidong}, \citenamefont {Rui}, \citenamefont {Son},
		\citenamefont {Kenji}, \citenamefont {Takashi}, \citenamefont {Fan},
		\citenamefont {Marc},\ and\ \citenamefont {Ning}}]{Emilio2022}%
	\BibitemOpen
	\bibfield  {author} {\bibinfo {author} {\bibfnamefont {C.}~\bibnamefont
			{Emilio}}, \bibinfo {author} {\bibfnamefont {W.}~\bibnamefont {Qiyue}},
		\bibinfo {author} {\bibfnamefont {K.}~\bibnamefont {Ryan}}, \bibinfo {author}
		{\bibfnamefont {C.}~\bibnamefont {Shi}}, \bibinfo {author} {\bibfnamefont
			{T.}~\bibnamefont {Haidong}}, \bibinfo {author} {\bibfnamefont
			{L.}~\bibnamefont {Rui}}, \bibinfo {author} {\bibfnamefont {T.}~\bibnamefont
			{Son}}, \bibinfo {author} {\bibfnamefont {W.}~\bibnamefont {Kenji}}, \bibinfo
		{author} {\bibfnamefont {T.}~\bibnamefont {Takashi}}, \bibinfo {author}
		{\bibfnamefont {Z.}~\bibnamefont {Fan}}, \bibinfo {author} {\bibfnamefont
			{B.}~\bibnamefont {Marc}},\ and\ \bibinfo {author} {\bibfnamefont {L.~C.}\
			\bibnamefont {Ning}},\ }\href {https://doi.org/10.1126/sciadv.aaw9770}
	{\bibfield  {journal} {\bibinfo  {journal} {Sci. Adv.}\ }\textbf {\bibinfo
			{volume} {5}},\ \bibinfo {pages} {eaaw9770} (\bibinfo {year}
		{2022})}\BibitemShut {NoStop}%
	\bibitem [{\citenamefont {Arora}\ \emph {et~al.}(2020)\citenamefont {Arora},
		\citenamefont {Polski}, \citenamefont {Zhang}, \citenamefont {Thomson},
		\citenamefont {Choi}, \citenamefont {Kim}, \citenamefont {Lin}, \citenamefont
		{Wilson}, \citenamefont {Xu}, \citenamefont {Chu}, \citenamefont {Watanabe},
		\citenamefont {Taniguchi}, \citenamefont {Alicea},\ and\ \citenamefont
		{Nadj-Perge}}]{Arora2020}%
	\BibitemOpen
	\bibfield  {author} {\bibinfo {author} {\bibfnamefont {H.~S.}\ \bibnamefont
			{Arora}}, \bibinfo {author} {\bibfnamefont {R.}~\bibnamefont {Polski}},
		\bibinfo {author} {\bibfnamefont {Y.}~\bibnamefont {Zhang}}, \bibinfo
		{author} {\bibfnamefont {A.}~\bibnamefont {Thomson}}, \bibinfo {author}
		{\bibfnamefont {Y.}~\bibnamefont {Choi}}, \bibinfo {author} {\bibfnamefont
			{H.}~\bibnamefont {Kim}}, \bibinfo {author} {\bibfnamefont {Z.}~\bibnamefont
			{Lin}}, \bibinfo {author} {\bibfnamefont {I.~Z.}\ \bibnamefont {Wilson}},
		\bibinfo {author} {\bibfnamefont {X.}~\bibnamefont {Xu}}, \bibinfo {author}
		{\bibfnamefont {J.-H.}\ \bibnamefont {Chu}}, \bibinfo {author} {\bibfnamefont
			{K.}~\bibnamefont {Watanabe}}, \bibinfo {author} {\bibfnamefont
			{T.}~\bibnamefont {Taniguchi}}, \bibinfo {author} {\bibfnamefont
			{J.}~\bibnamefont {Alicea}},\ and\ \bibinfo {author} {\bibfnamefont
			{S.}~\bibnamefont {Nadj-Perge}},\ }\href
	{https://doi.org/10.1038/s41586-020-2473-8} {\bibfield  {journal} {\bibinfo
			{journal} {Nature}\ }\textbf {\bibinfo {volume} {583}},\ \bibinfo {pages}
		{379} (\bibinfo {year} {2020})}\BibitemShut {NoStop}%
	\bibitem [{\citenamefont {Dagotto}(1994)}]{Dagotto1994}%
	\BibitemOpen
	\bibfield  {author} {\bibinfo {author} {\bibfnamefont {E.}~\bibnamefont
			{Dagotto}},\ }\href {https://doi.org/10.1103/RevModPhys.66.763} {\bibfield
		{journal} {\bibinfo  {journal} {Rev. Mod. Phys.}\ }\textbf {\bibinfo {volume}
			{66}},\ \bibinfo {pages} {763} (\bibinfo {year} {1994})}\BibitemShut
	{NoStop}%
	\bibitem [{\citenamefont {Imada}\ \emph {et~al.}(1998)\citenamefont {Imada},
		\citenamefont {Fujimori},\ and\ \citenamefont {Tokura}}]{Imada1998}%
	\BibitemOpen
	\bibfield  {author} {\bibinfo {author} {\bibfnamefont {M.}~\bibnamefont
			{Imada}}, \bibinfo {author} {\bibfnamefont {A.}~\bibnamefont {Fujimori}},\
		and\ \bibinfo {author} {\bibfnamefont {Y.}~\bibnamefont {Tokura}},\ }\href
	{https://doi.org/10.1103/RevModPhys.70.1039} {\bibfield  {journal} {\bibinfo
			{journal} {Rev. Mod. Phys.}\ }\textbf {\bibinfo {volume} {70}},\ \bibinfo
		{pages} {1039} (\bibinfo {year} {1998})}\BibitemShut {NoStop}%
	\bibitem [{\citenamefont {Orenstein}\ and\ \citenamefont
		{Millis}(2000)}]{Orenstein2000}%
	\BibitemOpen
	\bibfield  {author} {\bibinfo {author} {\bibfnamefont {J.}~\bibnamefont
			{Orenstein}}\ and\ \bibinfo {author} {\bibfnamefont {A.~J.}\ \bibnamefont
			{Millis}},\ }\href {https://doi.org/10.1126/science.288.5465.468} {\bibfield
		{journal} {\bibinfo  {journal} {Science}\ }\textbf {\bibinfo {volume}
			{288}},\ \bibinfo {pages} {468} (\bibinfo {year} {2000})}\BibitemShut
	{NoStop}%
	\bibitem [{\citenamefont {Lee}\ \emph {et~al.}(2006)\citenamefont {Lee},
		\citenamefont {Nagaosa},\ and\ \citenamefont {Wen}}]{Lee2006}%
	\BibitemOpen
	\bibfield  {author} {\bibinfo {author} {\bibfnamefont {P.~A.}\ \bibnamefont
			{Lee}}, \bibinfo {author} {\bibfnamefont {N.}~\bibnamefont {Nagaosa}},\ and\
		\bibinfo {author} {\bibfnamefont {X.-G.}\ \bibnamefont {Wen}},\ }\href
	{https://doi.org/10.1103/RevModPhys.78.17} {\bibfield  {journal} {\bibinfo
			{journal} {Rev. Mod. Phys.}\ }\textbf {\bibinfo {volume} {78}},\ \bibinfo
		{pages} {17} (\bibinfo {year} {2006})}\BibitemShut {NoStop}%
	\bibitem [{\citenamefont {Scalapino}(2012)}]{Scalapino2012}%
	\BibitemOpen
	\bibfield  {author} {\bibinfo {author} {\bibfnamefont {D.~J.}\ \bibnamefont
			{Scalapino}},\ }\href {https://doi.org/10.1103/RevModPhys.84.1383} {\bibfield
		{journal} {\bibinfo  {journal} {Rev. Mod. Phys.}\ }\textbf {\bibinfo
			{volume} {84}},\ \bibinfo {pages} {1383} (\bibinfo {year}
		{2012})}\BibitemShut {NoStop}%
	\bibitem [{\citenamefont {Fradkin}\ \emph {et~al.}(2015)\citenamefont
		{Fradkin}, \citenamefont {Kivelson},\ and\ \citenamefont
		{Tranquada}}]{Fradkin2015}%
	\BibitemOpen
	\bibfield  {author} {\bibinfo {author} {\bibfnamefont {E.}~\bibnamefont
			{Fradkin}}, \bibinfo {author} {\bibfnamefont {S.~A.}\ \bibnamefont
			{Kivelson}},\ and\ \bibinfo {author} {\bibfnamefont {J.~M.}\ \bibnamefont
			{Tranquada}},\ }\href {https://doi.org/10.1103/RevModPhys.87.457} {\bibfield
		{journal} {\bibinfo  {journal} {Rev. Mod. Phys.}\ }\textbf {\bibinfo {volume}
			{87}},\ \bibinfo {pages} {457} (\bibinfo {year} {2015})}\BibitemShut
	{NoStop}%
	\bibitem [{\citenamefont {Keimer}\ \emph {et~al.}(2015)\citenamefont {Keimer},
		\citenamefont {Kivelson}, \citenamefont {Norman}, \citenamefont {Uchida},\
		and\ \citenamefont {Zaanen}}]{Keimer2015}%
	\BibitemOpen
	\bibfield  {author} {\bibinfo {author} {\bibfnamefont {B.}~\bibnamefont
			{Keimer}}, \bibinfo {author} {\bibfnamefont {S.~A.}\ \bibnamefont
			{Kivelson}}, \bibinfo {author} {\bibfnamefont {M.~R.}\ \bibnamefont
			{Norman}}, \bibinfo {author} {\bibfnamefont {S.}~\bibnamefont {Uchida}},\
		and\ \bibinfo {author} {\bibfnamefont {J.}~\bibnamefont {Zaanen}},\ }\href
	{https://doi.org/10.1038/nature14165} {\bibfield  {journal} {\bibinfo
			{journal} {Nature}\ }\textbf {\bibinfo {volume} {518}},\ \bibinfo {pages}
		{179} (\bibinfo {year} {2015})}\BibitemShut {NoStop}%
	\bibitem [{\citenamefont {Batten}\ \emph {et~al.}(2013)\citenamefont {Batten},
		\citenamefont {Champness}, \citenamefont {Chen}, \citenamefont
		{Garcia-Martinez}, \citenamefont {Kitagawa}, \citenamefont {Öhrström},
		\citenamefont {O’Keeffe}, \citenamefont {Suh},\ and\ \citenamefont
		{Reedijk}}]{Batten2013}%
	\BibitemOpen
	\bibfield  {author} {\bibinfo {author} {\bibfnamefont {S.~R.}\ \bibnamefont
			{Batten}}, \bibinfo {author} {\bibfnamefont {N.~R.}\ \bibnamefont
			{Champness}}, \bibinfo {author} {\bibfnamefont {X.-M.}\ \bibnamefont {Chen}},
		\bibinfo {author} {\bibfnamefont {J.}~\bibnamefont {Garcia-Martinez}},
		\bibinfo {author} {\bibfnamefont {S.}~\bibnamefont {Kitagawa}}, \bibinfo
		{author} {\bibfnamefont {L.}~\bibnamefont {Öhrström}}, \bibinfo {author}
		{\bibfnamefont {M.}~\bibnamefont {O’Keeffe}}, \bibinfo {author}
		{\bibfnamefont {M.~P.}\ \bibnamefont {Suh}},\ and\ \bibinfo {author}
		{\bibfnamefont {J.}~\bibnamefont {Reedijk}},\ }\href
	{https://doi.org/doi:10.1351/PAC-REC-12-11-20} {\bibfield  {journal}
		{\bibinfo  {journal} {Pure Appl. Chem.}\ }\textbf {\bibinfo {volume} {85}},\
		\bibinfo {pages} {1715} (\bibinfo {year} {2013})}\BibitemShut {NoStop}%
	\bibitem [{\citenamefont {Murase}\ \emph
		{et~al.}(2017{\natexlab{a}})\citenamefont {Murase}, \citenamefont {Leong},\
		and\ \citenamefont {D’Alessandro}}]{Murase2017}%
	\BibitemOpen
	\bibfield  {author} {\bibinfo {author} {\bibfnamefont {R.}~\bibnamefont
			{Murase}}, \bibinfo {author} {\bibfnamefont {C.~F.}\ \bibnamefont {Leong}},\
		and\ \bibinfo {author} {\bibfnamefont {D.~M.}\ \bibnamefont
			{D’Alessandro}},\ }\href {https://doi.org/10.1021/acs.inorgchem.7b02090}
	{\bibfield  {journal} {\bibinfo  {journal} {Inorg. Chem.}\ }\textbf {\bibinfo
			{volume} {56}},\ \bibinfo {pages} {14373} (\bibinfo {year}
		{2017}{\natexlab{a}})}\BibitemShut {NoStop}%
	\bibitem [{\citenamefont {Murase}\ \emph
		{et~al.}(2017{\natexlab{b}})\citenamefont {Murase}, \citenamefont {Abrahams},
		\citenamefont {D’Alessandro}, \citenamefont {Davies}, \citenamefont
		{Hudson}, \citenamefont {Jameson}, \citenamefont {Moubaraki}, \citenamefont
		{Murray}, \citenamefont {Robson},\ and\ \citenamefont
		{Sutton}}]{Murase2017b}%
	\BibitemOpen
	\bibfield  {author} {\bibinfo {author} {\bibfnamefont {R.}~\bibnamefont
			{Murase}}, \bibinfo {author} {\bibfnamefont {B.~F.}\ \bibnamefont
			{Abrahams}}, \bibinfo {author} {\bibfnamefont {D.~M.}\ \bibnamefont
			{D’Alessandro}}, \bibinfo {author} {\bibfnamefont {C.~G.}\ \bibnamefont
			{Davies}}, \bibinfo {author} {\bibfnamefont {T.~A.}\ \bibnamefont {Hudson}},
		\bibinfo {author} {\bibfnamefont {G.~N.~L.}\ \bibnamefont {Jameson}},
		\bibinfo {author} {\bibfnamefont {B.}~\bibnamefont {Moubaraki}}, \bibinfo
		{author} {\bibfnamefont {K.~S.}\ \bibnamefont {Murray}}, \bibinfo {author}
		{\bibfnamefont {R.}~\bibnamefont {Robson}},\ and\ \bibinfo {author}
		{\bibfnamefont {A.~L.}\ \bibnamefont {Sutton}},\ }\href
	{https://doi.org/10.1021/acs.inorgchem.7b01038} {\bibfield  {journal}
		{\bibinfo  {journal} {Inorg. Chem.}\ }\textbf {\bibinfo {volume} {56}},\
		\bibinfo {pages} {9025} (\bibinfo {year} {2017}{\natexlab{b}})}\BibitemShut
	{NoStop}%
	\bibitem [{\citenamefont {Kingsbury}\ \emph {et~al.}(2017)\citenamefont
		{Kingsbury}, \citenamefont {Abrahams}, \citenamefont {D’Alessandro},
		\citenamefont {Hudson}, \citenamefont {Murase}, \citenamefont {Robson},\ and\
		\citenamefont {White}}]{Kingsbury2017}%
	\BibitemOpen
	\bibfield  {author} {\bibinfo {author} {\bibfnamefont {C.~J.}\ \bibnamefont
			{Kingsbury}}, \bibinfo {author} {\bibfnamefont {B.~F.}\ \bibnamefont
			{Abrahams}}, \bibinfo {author} {\bibfnamefont {D.~M.}\ \bibnamefont
			{D’Alessandro}}, \bibinfo {author} {\bibfnamefont {T.~A.}\ \bibnamefont
			{Hudson}}, \bibinfo {author} {\bibfnamefont {R.}~\bibnamefont {Murase}},
		\bibinfo {author} {\bibfnamefont {R.}~\bibnamefont {Robson}},\ and\ \bibinfo
		{author} {\bibfnamefont {K.~F.}\ \bibnamefont {White}},\ }\href
	{https://doi.org/10.1021/acs.cgd.6b01886} {\bibfield  {journal} {\bibinfo
			{journal} {Cryst. Growth Des.}\ }\textbf {\bibinfo {volume} {17}},\ \bibinfo
		{pages} {1465} (\bibinfo {year} {2017})}\BibitemShut {NoStop}%
	\bibitem [{\citenamefont {Jeon}\ \emph {et~al.}(2015)\citenamefont {Jeon},
		\citenamefont {Negru}, \citenamefont {Van~Duyne},\ and\ \citenamefont
		{Harris}}]{Jeon2015}%
	\BibitemOpen
	\bibfield  {author} {\bibinfo {author} {\bibfnamefont {I.-R.}\ \bibnamefont
			{Jeon}}, \bibinfo {author} {\bibfnamefont {B.}~\bibnamefont {Negru}},
		\bibinfo {author} {\bibfnamefont {R.~P.}\ \bibnamefont {Van~Duyne}},\ and\
		\bibinfo {author} {\bibfnamefont {T.~D.}\ \bibnamefont {Harris}},\ }\href
	{https://doi.org/10.1021/jacs.5b10382} {\bibfield  {journal} {\bibinfo
			{journal} {J. Am. Chem. Soc.}\ }\textbf {\bibinfo {volume} {137}},\ \bibinfo
		{pages} {15699} (\bibinfo {year} {2015})}\BibitemShut {NoStop}%
	\bibitem [{\citenamefont {Darago}\ \emph {et~al.}(2015)\citenamefont {Darago},
		\citenamefont {Aubrey}, \citenamefont {Yu}, \citenamefont {Gonzalez},\ and\
		\citenamefont {Long}}]{Darago2015}%
	\BibitemOpen
	\bibfield  {author} {\bibinfo {author} {\bibfnamefont {L.~E.}\ \bibnamefont
			{Darago}}, \bibinfo {author} {\bibfnamefont {M.~L.}\ \bibnamefont {Aubrey}},
		\bibinfo {author} {\bibfnamefont {C.~J.}\ \bibnamefont {Yu}}, \bibinfo
		{author} {\bibfnamefont {M.~I.}\ \bibnamefont {Gonzalez}},\ and\ \bibinfo
		{author} {\bibfnamefont {J.~R.}\ \bibnamefont {Long}},\ }\href
	{https://doi.org/10.1021/jacs.5b10385} {\bibfield  {journal} {\bibinfo
			{journal} {J. Am. Chem. Soc.}\ }\textbf {\bibinfo {volume} {137}},\ \bibinfo
		{pages} {15703} (\bibinfo {year} {2015})}\BibitemShut {NoStop}%
	\bibitem [{\citenamefont {DeGayner}\ \emph {et~al.}(2017)\citenamefont
		{DeGayner}, \citenamefont {Jeon}, \citenamefont {Sun}, \citenamefont
		{Dincă},\ and\ \citenamefont {Harris}}]{DeGayner2017}%
	\BibitemOpen
	\bibfield  {author} {\bibinfo {author} {\bibfnamefont {J.~A.}\ \bibnamefont
			{DeGayner}}, \bibinfo {author} {\bibfnamefont {I.-R.}\ \bibnamefont {Jeon}},
		\bibinfo {author} {\bibfnamefont {L.}~\bibnamefont {Sun}}, \bibinfo {author}
		{\bibfnamefont {M.}~\bibnamefont {Dincă}},\ and\ \bibinfo {author}
		{\bibfnamefont {T.~D.}\ \bibnamefont {Harris}},\ }\href
	{https://doi.org/10.1021/jacs.7b00705} {\bibfield  {journal} {\bibinfo
			{journal} {J. Am. Chem. Soc.}\ }\textbf {\bibinfo {volume} {139}},\ \bibinfo
		{pages} {4175} (\bibinfo {year} {2017})}\BibitemShut {NoStop}%
	\bibitem [{\citenamefont {Henling}\ and\ \citenamefont
		{Marsh}(2014)}]{Henling2014}%
	\BibitemOpen
	\bibfield  {author} {\bibinfo {author} {\bibfnamefont {L.~M.}\ \bibnamefont
			{Henling}}\ and\ \bibinfo {author} {\bibfnamefont {R.~E.}\ \bibnamefont
			{Marsh}},\ }\href {https://doi.org/10.1107/S2053229614017549} {\bibfield
		{journal} {\bibinfo  {journal} {Acta Crystallogr., Sect. C: Struct. Chem.}\
		}\textbf {\bibinfo {volume} {70}},\ \bibinfo {pages} {834} (\bibinfo {year}
		{2014})},\ \bibinfo {note} {{C}SD-FAZGIY}\BibitemShut {NoStop}%
	\bibitem [{\citenamefont {Henline}\ \emph {et~al.}(2014)\citenamefont
		{Henline}, \citenamefont {Wang}, \citenamefont {Pike}, \citenamefont {Ahern},
		\citenamefont {Sousa}, \citenamefont {Patterson}, \citenamefont {Kerr},\ and\
		\citenamefont {Cahill}}]{Henline2014}%
	\BibitemOpen
	\bibfield  {author} {\bibinfo {author} {\bibfnamefont {K.~M.}\ \bibnamefont
			{Henline}}, \bibinfo {author} {\bibfnamefont {C.}~\bibnamefont {Wang}},
		\bibinfo {author} {\bibfnamefont {R.~D.}\ \bibnamefont {Pike}}, \bibinfo
		{author} {\bibfnamefont {J.~C.}\ \bibnamefont {Ahern}}, \bibinfo {author}
		{\bibfnamefont {B.}~\bibnamefont {Sousa}}, \bibinfo {author} {\bibfnamefont
			{H.~H.}\ \bibnamefont {Patterson}}, \bibinfo {author} {\bibfnamefont {A.~T.}\
			\bibnamefont {Kerr}},\ and\ \bibinfo {author} {\bibfnamefont {C.~L.}\
			\bibnamefont {Cahill}},\ }\href {https://doi.org/10.1021/cg500005p}
	{\bibfield  {journal} {\bibinfo  {journal} {Cryst. Growth Des.}\ }\textbf
		{\bibinfo {volume} {14}},\ \bibinfo {pages} {1449} (\bibinfo {year}
		{2014})}\BibitemShut {NoStop}%
	\bibitem [{\citenamefont {Polunin}\ \emph {et~al.}(2015)\citenamefont
		{Polunin}, \citenamefont {Dorofeeva}, \citenamefont {Baranchikov},
		\citenamefont {Ivanov}, \citenamefont {Gavrilenko}, \citenamefont {Kiskin},
		\citenamefont {Eremenko}, \citenamefont {Novotortsev},\ and\ \citenamefont
		{Kolotilov}}]{Polunin2015}%
	\BibitemOpen
	\bibfield  {author} {\bibinfo {author} {\bibfnamefont {R.~A.}\ \bibnamefont
			{Polunin}}, \bibinfo {author} {\bibfnamefont {V.~N.}\ \bibnamefont
			{Dorofeeva}}, \bibinfo {author} {\bibfnamefont {A.~E.}\ \bibnamefont
			{Baranchikov}}, \bibinfo {author} {\bibfnamefont {V.~K.}\ \bibnamefont
			{Ivanov}}, \bibinfo {author} {\bibfnamefont {K.~S.}\ \bibnamefont
			{Gavrilenko}}, \bibinfo {author} {\bibfnamefont {M.~A.}\ \bibnamefont
			{Kiskin}}, \bibinfo {author} {\bibfnamefont {I.~L.}\ \bibnamefont
			{Eremenko}}, \bibinfo {author} {\bibfnamefont {V.~M.}\ \bibnamefont
			{Novotortsev}},\ and\ \bibinfo {author} {\bibfnamefont {S.~V.}\ \bibnamefont
			{Kolotilov}},\ }\href {https://doi.org/10.1134/S1070328415060056} {\bibfield
		{journal} {\bibinfo  {journal} {Russ. J. Coord. Chem.}\ }\textbf {\bibinfo
			{volume} {41}},\ \bibinfo {pages} {353} (\bibinfo {year} {2015})}\BibitemShut
	{NoStop}%
	\bibitem [{\citenamefont {Kalmutzki}\ \emph {et~al.}(2018)\citenamefont
		{Kalmutzki}, \citenamefont {Hanikel},\ and\ \citenamefont
		{Yaghi}}]{Kalmutzki2018}%
	\BibitemOpen
	\bibfield  {author} {\bibinfo {author} {\bibfnamefont {M.~J.}\ \bibnamefont
			{Kalmutzki}}, \bibinfo {author} {\bibfnamefont {N.}~\bibnamefont {Hanikel}},\
		and\ \bibinfo {author} {\bibfnamefont {O.~M.}\ \bibnamefont {Yaghi}},\ }\href
	{https://doi.org/10.1126/sciadv.aat9180} {\bibfield  {journal} {\bibinfo
			{journal} {Sci. Adv.}\ }\textbf {\bibinfo {volume} {4}},\ \bibinfo {pages}
		{eaat9180} (\bibinfo {year} {2018})}\BibitemShut {NoStop}%
	\bibitem [{\citenamefont {Jiang}\ \emph {et~al.}(2019)\citenamefont {Jiang},
		\citenamefont {Liu}, \citenamefont {Mei}, \citenamefont {Cui},\ and\
		\citenamefont {Liu}}]{Jiang2019}%
	\BibitemOpen
	\bibfield  {author} {\bibinfo {author} {\bibfnamefont {W.}~\bibnamefont
			{Jiang}}, \bibinfo {author} {\bibfnamefont {Z.}~\bibnamefont {Liu}}, \bibinfo
		{author} {\bibfnamefont {J.-W.}\ \bibnamefont {Mei}}, \bibinfo {author}
		{\bibfnamefont {B.}~\bibnamefont {Cui}},\ and\ \bibinfo {author}
		{\bibfnamefont {F.}~\bibnamefont {Liu}},\ }\href
	{https://doi.org/10.1039/C8NR08479C} {\bibfield  {journal} {\bibinfo
			{journal} {Nanoscale}\ }\textbf {\bibinfo {volume} {11}},\ \bibinfo {pages}
		{955} (\bibinfo {year} {2019})}\BibitemShut {NoStop}%
	\bibitem [{\citenamefont {Kumar}\ \emph {et~al.}(2021)\citenamefont {Kumar},
		\citenamefont {Hellerstedt}, \citenamefont {Field}, \citenamefont {Lowe},
		\citenamefont {Yin}, \citenamefont {Medhekar},\ and\ \citenamefont
		{Schiffrin}}]{Kumar2021}%
	\BibitemOpen
	\bibfield  {author} {\bibinfo {author} {\bibfnamefont {D.}~\bibnamefont
			{Kumar}}, \bibinfo {author} {\bibfnamefont {J.}~\bibnamefont {Hellerstedt}},
		\bibinfo {author} {\bibfnamefont {B.}~\bibnamefont {Field}}, \bibinfo
		{author} {\bibfnamefont {B.}~\bibnamefont {Lowe}}, \bibinfo {author}
		{\bibfnamefont {Y.}~\bibnamefont {Yin}}, \bibinfo {author} {\bibfnamefont
			{N.~V.}\ \bibnamefont {Medhekar}},\ and\ \bibinfo {author} {\bibfnamefont
			{A.}~\bibnamefont {Schiffrin}},\ }\href
	{https://doi.org/https://doi.org/10.1002/adfm.202106474} {\bibfield
		{journal} {\bibinfo  {journal} {Adv. Funct. Mater}\ }\textbf {\bibinfo
			{volume} {31}},\ \bibinfo {pages} {2106474} (\bibinfo {year}
		{2021})}\BibitemShut {NoStop}%
	\bibitem [{\citenamefont {Zhang}\ \emph {et~al.}(2017)\citenamefont {Zhang},
		\citenamefont {Zhou}, \citenamefont {Cui}, \citenamefont {Zhao},\ and\
		\citenamefont {Liu}}]{Zhang2017}%
	\BibitemOpen
	\bibfield  {author} {\bibinfo {author} {\bibfnamefont {X.}~\bibnamefont
			{Zhang}}, \bibinfo {author} {\bibfnamefont {Y.}~\bibnamefont {Zhou}},
		\bibinfo {author} {\bibfnamefont {B.}~\bibnamefont {Cui}}, \bibinfo {author}
		{\bibfnamefont {M.}~\bibnamefont {Zhao}},\ and\ \bibinfo {author}
		{\bibfnamefont {F.}~\bibnamefont {Liu}},\ }\href
	{https://doi.org/10.1021/acs.nanolett.7b02795} {\bibfield  {journal}
		{\bibinfo  {journal} {Nano Lett.}\ }\textbf {\bibinfo {volume} {17}},\
		\bibinfo {pages} {6166} (\bibinfo {year} {2017})}\BibitemShut {NoStop}%
	\bibitem [{\citenamefont {Huang}\ \emph {et~al.}(2018)\citenamefont {Huang},
		\citenamefont {Zhang}, \citenamefont {Liu}, \citenamefont {Yu}, \citenamefont
		{Chen}, \citenamefont {Xu},\ and\ \citenamefont {Zhu}}]{Huang2018}%
	\BibitemOpen
	\bibfield  {author} {\bibinfo {author} {\bibfnamefont {X.}~\bibnamefont
			{Huang}}, \bibinfo {author} {\bibfnamefont {S.}~\bibnamefont {Zhang}},
		\bibinfo {author} {\bibfnamefont {L.}~\bibnamefont {Liu}}, \bibinfo {author}
		{\bibfnamefont {L.}~\bibnamefont {Yu}}, \bibinfo {author} {\bibfnamefont
			{G.}~\bibnamefont {Chen}}, \bibinfo {author} {\bibfnamefont {W.}~\bibnamefont
			{Xu}},\ and\ \bibinfo {author} {\bibfnamefont {D.}~\bibnamefont {Zhu}},\
	}\href {https://doi.org/https://doi.org/10.1002/anie.201707568} {\bibfield
		{journal} {\bibinfo  {journal} {Angew. Chem.}\ }\textbf {\bibinfo {volume}
			{57}},\ \bibinfo {pages} {146} (\bibinfo {year} {2018})}\BibitemShut
	{NoStop}%
	\bibitem [{\citenamefont {Takenaka}\ \emph {et~al.}(2021)\citenamefont
		{Takenaka}, \citenamefont {Ishihara}, \citenamefont {Roppongi}, \citenamefont
		{Miao}, \citenamefont {Mizukami}, \citenamefont {Makita}, \citenamefont
		{Tsurumi}, \citenamefont {Watanabe}, \citenamefont {Takeya}, \citenamefont
		{Yamashita}, \citenamefont {Torizuka}, \citenamefont {Uwatoko}, \citenamefont
		{Sasaki}, \citenamefont {Huang}, \citenamefont {Xu}, \citenamefont {Zhu},
		\citenamefont {Su}, \citenamefont {Cheng}, \citenamefont {Shibauchi},\ and\
		\citenamefont {Hashimoto}}]{Takenaka2021}%
	\BibitemOpen
	\bibfield  {author} {\bibinfo {author} {\bibfnamefont {T.}~\bibnamefont
			{Takenaka}}, \bibinfo {author} {\bibfnamefont {K.}~\bibnamefont {Ishihara}},
		\bibinfo {author} {\bibfnamefont {M.}~\bibnamefont {Roppongi}}, \bibinfo
		{author} {\bibfnamefont {Y.}~\bibnamefont {Miao}}, \bibinfo {author}
		{\bibfnamefont {Y.}~\bibnamefont {Mizukami}}, \bibinfo {author}
		{\bibfnamefont {T.}~\bibnamefont {Makita}}, \bibinfo {author} {\bibfnamefont
			{J.}~\bibnamefont {Tsurumi}}, \bibinfo {author} {\bibfnamefont
			{S.}~\bibnamefont {Watanabe}}, \bibinfo {author} {\bibfnamefont
			{J.}~\bibnamefont {Takeya}}, \bibinfo {author} {\bibfnamefont
			{M.}~\bibnamefont {Yamashita}}, \bibinfo {author} {\bibfnamefont
			{K.}~\bibnamefont {Torizuka}}, \bibinfo {author} {\bibfnamefont
			{Y.}~\bibnamefont {Uwatoko}}, \bibinfo {author} {\bibfnamefont
			{T.}~\bibnamefont {Sasaki}}, \bibinfo {author} {\bibfnamefont
			{X.}~\bibnamefont {Huang}}, \bibinfo {author} {\bibfnamefont
			{W.}~\bibnamefont {Xu}}, \bibinfo {author} {\bibfnamefont {D.}~\bibnamefont
			{Zhu}}, \bibinfo {author} {\bibfnamefont {N.}~\bibnamefont {Su}}, \bibinfo
		{author} {\bibfnamefont {J.-G.}\ \bibnamefont {Cheng}}, \bibinfo {author}
		{\bibfnamefont {T.}~\bibnamefont {Shibauchi}},\ and\ \bibinfo {author}
		{\bibfnamefont {K.}~\bibnamefont {Hashimoto}},\ }\href
	{https://doi.org/10.1126/sciadv.abf3996} {\bibfield  {journal} {\bibinfo
			{journal} {Sci. Adv.}\ }\textbf {\bibinfo {volume} {7}},\ \bibinfo {pages}
		{eabf3996} (\bibinfo {year} {2021})}\BibitemShut {NoStop}%
	\bibitem [{\citenamefont {Yaghi}(2016)}]{Yaghi2016}%
	\BibitemOpen
	\bibfield  {author} {\bibinfo {author} {\bibfnamefont {O.~M.}\ \bibnamefont
			{Yaghi}},\ }\href {https://doi.org/10.1021/jacs.6b11821} {\bibfield
		{journal} {\bibinfo  {journal} {J. Am. Chem. Soc.}\ }\textbf {\bibinfo
			{volume} {138}},\ \bibinfo {pages} {15507} (\bibinfo {year}
		{2016})}\BibitemShut {NoStop}%
	\bibitem [{\citenamefont {Nourse}\ \emph
		{et~al.}(2021{\natexlab{a}})\citenamefont {Nourse}, \citenamefont
		{McKenzie},\ and\ \citenamefont {Powell}}]{Nourse2021a}%
	\BibitemOpen
	\bibfield  {author} {\bibinfo {author} {\bibfnamefont {H.~L.}\ \bibnamefont
			{Nourse}}, \bibinfo {author} {\bibfnamefont {R.~H.}\ \bibnamefont
			{McKenzie}},\ and\ \bibinfo {author} {\bibfnamefont {B.~J.}\ \bibnamefont
			{Powell}},\ }\href {https://doi.org/10.1103/PhysRevB.103.L081114} {\bibfield
		{journal} {\bibinfo  {journal} {Phys. Rev. B}\ }\textbf {\bibinfo {volume}
			{103}},\ \bibinfo {pages} {L081114} (\bibinfo {year}
		{2021}{\natexlab{a}})}\BibitemShut {NoStop}%
	\bibitem [{\citenamefont {Nourse}\ \emph
		{et~al.}(2021{\natexlab{b}})\citenamefont {Nourse}, \citenamefont
		{McKenzie},\ and\ \citenamefont {Powell}}]{Nourse2021b}%
	\BibitemOpen
	\bibfield  {author} {\bibinfo {author} {\bibfnamefont {H.~L.}\ \bibnamefont
			{Nourse}}, \bibinfo {author} {\bibfnamefont {R.~H.}\ \bibnamefont
			{McKenzie}},\ and\ \bibinfo {author} {\bibfnamefont {B.~J.}\ \bibnamefont
			{Powell}},\ }\href {https://doi.org/10.1103/PhysRevB.104.075104} {\bibfield
		{journal} {\bibinfo  {journal} {Phys. Rev. B}\ }\textbf {\bibinfo {volume}
			{104}},\ \bibinfo {pages} {075104} (\bibinfo {year}
		{2021}{\natexlab{b}})}\BibitemShut {NoStop}%
	\bibitem [{\citenamefont {Merino}\ \emph {et~al.}(2021)\citenamefont {Merino},
		\citenamefont {L\'opez},\ and\ \citenamefont {Powell}}]{Merino2021}%
	\BibitemOpen
	\bibfield  {author} {\bibinfo {author} {\bibfnamefont {J.}~\bibnamefont
			{Merino}}, \bibinfo {author} {\bibfnamefont {M.~F.}\ \bibnamefont
			{L\'opez}},\ and\ \bibinfo {author} {\bibfnamefont {B.~J.}\ \bibnamefont
			{Powell}},\ }\href {https://doi.org/10.1103/PhysRevB.103.094517} {\bibfield
		{journal} {\bibinfo  {journal} {Phys. Rev. B}\ }\textbf {\bibinfo {volume}
			{103}},\ \bibinfo {pages} {094517} (\bibinfo {year} {2021})}\BibitemShut
	{NoStop}%
	\bibitem [{\citenamefont {Richter}\ \emph {et~al.}(2004)\citenamefont
		{Richter}, \citenamefont {Schulenburg}, \citenamefont {Honecker},\ and\
		\citenamefont {Schmalfu{\ss}}}]{Richter2004}%
	\BibitemOpen
	\bibfield  {author} {\bibinfo {author} {\bibfnamefont {J.}~\bibnamefont
			{Richter}}, \bibinfo {author} {\bibfnamefont {J.}~\bibnamefont
			{Schulenburg}}, \bibinfo {author} {\bibfnamefont {A.}~\bibnamefont
			{Honecker}},\ and\ \bibinfo {author} {\bibfnamefont {D.}~\bibnamefont
			{Schmalfu{\ss}}},\ }\href {https://doi.org/10.1103/PhysRevB.70.174454}
	{\bibfield  {journal} {\bibinfo  {journal} {Phys. Rev. B}\ }\textbf {\bibinfo
			{volume} {70}},\ \bibinfo {pages} {174454} (\bibinfo {year}
		{2004})}\BibitemShut {NoStop}%
	\bibitem [{\citenamefont {Misguich}\ and\ \citenamefont
		{Sindzingre}(2007)}]{Misguich2007}%
	\BibitemOpen
	\bibfield  {author} {\bibinfo {author} {\bibfnamefont {G.}~\bibnamefont
			{Misguich}}\ and\ \bibinfo {author} {\bibfnamefont {P.}~\bibnamefont
			{Sindzingre}},\ }\href {https://doi.org/10.1088/0953-8984/19/14/145202}
	{\bibfield  {journal} {\bibinfo  {journal} {J. Phys.: Condens. Matter}\
		}\textbf {\bibinfo {volume} {19}},\ \bibinfo {pages} {145202} (\bibinfo
		{year} {2007})}\BibitemShut {NoStop}%
	\bibitem [{\citenamefont {Yang}\ \emph {et~al.}(2010)\citenamefont {Yang},
		\citenamefont {Paramekanti},\ and\ \citenamefont {Kim}}]{Yang2010}%
	\BibitemOpen
	\bibfield  {author} {\bibinfo {author} {\bibfnamefont {B.-J.}\ \bibnamefont
			{Yang}}, \bibinfo {author} {\bibfnamefont {A.}~\bibnamefont {Paramekanti}},\
		and\ \bibinfo {author} {\bibfnamefont {Y.~B.}\ \bibnamefont {Kim}},\ }\href
	{https://doi.org/10.1103/PhysRevB.81.134418} {\bibfield  {journal} {\bibinfo
			{journal} {Phys. Rev. B}\ }\textbf {\bibinfo {volume} {81}},\ \bibinfo
		{pages} {134418} (\bibinfo {year} {2010})}\BibitemShut {NoStop}%
	\bibitem [{\citenamefont {Jahromi}\ and\ \citenamefont
		{Or{\'{u}}s}(2018)}]{Jahromi2018}%
	\BibitemOpen
	\bibfield  {author} {\bibinfo {author} {\bibfnamefont {S.~S.}\ \bibnamefont
			{Jahromi}}\ and\ \bibinfo {author} {\bibfnamefont {R.}~\bibnamefont
			{Or{\'{u}}s}},\ }\href {https://doi.org/10.1103/PhysRevB.98.155108}
	{\bibfield  {journal} {\bibinfo  {journal} {Phys. Rev. B}\ }\textbf {\bibinfo
			{volume} {98}},\ \bibinfo {pages} {155108} (\bibinfo {year}
		{2018})}\BibitemShut {NoStop}%
	\bibitem [{\citenamefont {Bergman}\ \emph {et~al.}(2008)\citenamefont
		{Bergman}, \citenamefont {Wu},\ and\ \citenamefont {Balents}}]{Bergman2008}%
	\BibitemOpen
	\bibfield  {author} {\bibinfo {author} {\bibfnamefont {D.~L.}\ \bibnamefont
			{Bergman}}, \bibinfo {author} {\bibfnamefont {C.}~\bibnamefont {Wu}},\ and\
		\bibinfo {author} {\bibfnamefont {L.}~\bibnamefont {Balents}},\ }\href
	{https://doi.org/10.1103/PhysRevB.78.125104} {\bibfield  {journal} {\bibinfo
			{journal} {Phys. Rev. B}\ }\textbf {\bibinfo {volume} {78}},\ \bibinfo
		{pages} {125104} (\bibinfo {year} {2008})}\BibitemShut {NoStop}%
	\bibitem [{\citenamefont {Jacko}\ \emph {et~al.}(2015)\citenamefont {Jacko},
		\citenamefont {Janani}, \citenamefont {Koepernik},\ and\ \citenamefont
		{Powell}}]{Jacko2015}%
	\BibitemOpen
	\bibfield  {author} {\bibinfo {author} {\bibfnamefont {A.~C.}\ \bibnamefont
			{Jacko}}, \bibinfo {author} {\bibfnamefont {C.}~\bibnamefont {Janani}},
		\bibinfo {author} {\bibfnamefont {K.}~\bibnamefont {Koepernik}},\ and\
		\bibinfo {author} {\bibfnamefont {B.~J.}\ \bibnamefont {Powell}},\ }\href
	{http://link.aps.org/doi/10.1103/PhysRevB.91.125140} {\bibfield  {journal}
		{\bibinfo  {journal} {Phys. Rev. B}\ }\textbf {\bibinfo {volume} {91}},\
		\bibinfo {pages} {125140} (\bibinfo {year} {2015})}\BibitemShut {NoStop}%
	\bibitem [{\citenamefont {R\"uegg}\ \emph {et~al.}(2010)\citenamefont
		{R\"uegg}, \citenamefont {Wen},\ and\ \citenamefont {Fiete}}]{Ruegg2010}%
	\BibitemOpen
	\bibfield  {author} {\bibinfo {author} {\bibfnamefont {A.}~\bibnamefont
			{R\"uegg}}, \bibinfo {author} {\bibfnamefont {J.}~\bibnamefont {Wen}},\ and\
		\bibinfo {author} {\bibfnamefont {G.~A.}\ \bibnamefont {Fiete}},\ }\href
	{https://doi.org/10.1103/PhysRevB.81.205115} {\bibfield  {journal} {\bibinfo
			{journal} {Phys. Rev. B}\ }\textbf {\bibinfo {volume} {81}},\ \bibinfo
		{pages} {205115} (\bibinfo {year} {2010})}\BibitemShut {NoStop}%
	\bibitem [{\citenamefont {Kotliar}\ and\ \citenamefont
		{Ruckenstein}(1986)}]{Kotliar1986}%
	\BibitemOpen
	\bibfield  {author} {\bibinfo {author} {\bibfnamefont {G.}~\bibnamefont
			{Kotliar}}\ and\ \bibinfo {author} {\bibfnamefont {A.~E.}\ \bibnamefont
			{Ruckenstein}},\ }\href {http://link.aps.org/doi/10.1103/PhysRevLett.57.1362}
	{\bibfield  {journal} {\bibinfo  {journal} {Phys. Rev. Lett.}\ }\textbf
		{\bibinfo {volume} {57}},\ \bibinfo {pages} {1362} (\bibinfo {year}
		{1986})}\BibitemShut {NoStop}%
	\bibitem [{\citenamefont {Lechermann}\ \emph {et~al.}(2007)\citenamefont
		{Lechermann}, \citenamefont {Georges}, \citenamefont {Kotliar},\ and\
		\citenamefont {Parcollet}}]{Lechermann2007}%
	\BibitemOpen
	\bibfield  {author} {\bibinfo {author} {\bibfnamefont {F.}~\bibnamefont
			{Lechermann}}, \bibinfo {author} {\bibfnamefont {A.}~\bibnamefont {Georges}},
		\bibinfo {author} {\bibfnamefont {G.}~\bibnamefont {Kotliar}},\ and\ \bibinfo
		{author} {\bibfnamefont {O.}~\bibnamefont {Parcollet}},\ }\href
	{http://link.aps.org/doi/10.1103/PhysRevB.76.155102} {\bibfield  {journal}
		{\bibinfo  {journal} {Phys. Rev. B}\ }\textbf {\bibinfo {volume} {76}},\
		\bibinfo {pages} {155102} (\bibinfo {year} {2007})}\BibitemShut {NoStop}%
	\bibitem [{\citenamefont {Lanat{\`{a}}}\ \emph {et~al.}(2015)\citenamefont
		{Lanat{\`{a}}}, \citenamefont {Yao}, \citenamefont {Wang}, \citenamefont
		{Ho},\ and\ \citenamefont {Kotliar}}]{Lanata2015}%
	\BibitemOpen
	\bibfield  {author} {\bibinfo {author} {\bibfnamefont {N.}~\bibnamefont
			{Lanat{\`{a}}}}, \bibinfo {author} {\bibfnamefont {Y.}~\bibnamefont {Yao}},
		\bibinfo {author} {\bibfnamefont {C.-Z.}\ \bibnamefont {Wang}}, \bibinfo
		{author} {\bibfnamefont {K.-M.}\ \bibnamefont {Ho}},\ and\ \bibinfo {author}
		{\bibfnamefont {G.}~\bibnamefont {Kotliar}},\ }\href
	{http://link.aps.org/doi/10.1103/PhysRevX.5.011008} {\bibfield  {journal}
		{\bibinfo  {journal} {Phys. Rev. X}\ }\textbf {\bibinfo {volume} {5}},\
		\bibinfo {pages} {011008} (\bibinfo {year} {2015})}\BibitemShut {NoStop}%
	\bibitem [{\citenamefont {Lanat{\`{a}}}\ \emph {et~al.}(2017)\citenamefont
		{Lanat{\`{a}}}, \citenamefont {Yao}, \citenamefont {Deng}, \citenamefont
		{Dobrosavljevi{\'{c}}},\ and\ \citenamefont {Kotliar}}]{Lanata2017}%
	\BibitemOpen
	\bibfield  {author} {\bibinfo {author} {\bibfnamefont {N.}~\bibnamefont
			{Lanat{\`{a}}}}, \bibinfo {author} {\bibfnamefont {Y.}~\bibnamefont {Yao}},
		\bibinfo {author} {\bibfnamefont {X.}~\bibnamefont {Deng}}, \bibinfo {author}
		{\bibfnamefont {V.}~\bibnamefont {Dobrosavljevi{\'{c}}}},\ and\ \bibinfo
		{author} {\bibfnamefont {G.}~\bibnamefont {Kotliar}},\ }\href
	{https://link.aps.org/doi/10.1103/PhysRevLett.118.126401} {\bibfield
		{journal} {\bibinfo  {journal} {Phys. Rev. Lett.}\ }\textbf {\bibinfo
			{volume} {118}},\ \bibinfo {pages} {126401} (\bibinfo {year}
		{2017})}\BibitemShut {NoStop}%
	\bibitem [{\citenamefont {B\"unemann}\ and\ \citenamefont
		{Gebhard}(2007)}]{Bunemann2007}%
	\BibitemOpen
	\bibfield  {author} {\bibinfo {author} {\bibfnamefont {J.}~\bibnamefont
			{B\"unemann}}\ and\ \bibinfo {author} {\bibfnamefont {F.}~\bibnamefont
			{Gebhard}},\ }\href {https://doi.org/10.1103/PhysRevB.76.193104} {\bibfield
		{journal} {\bibinfo  {journal} {Phys. Rev. B}\ }\textbf {\bibinfo {volume}
			{76}},\ \bibinfo {pages} {193104} (\bibinfo {year} {2007})}\BibitemShut
	{NoStop}%
	\bibitem [{\citenamefont {Parcollet}\ \emph {et~al.}(2015)\citenamefont
		{Parcollet}, \citenamefont {Ferrero}, \citenamefont {Ayral}, \citenamefont
		{Hafermann}, \citenamefont {Krivenko}, \citenamefont {Messio},\ and\
		\citenamefont {Seth}}]{Parcollet2015}%
	\BibitemOpen
	\bibfield  {author} {\bibinfo {author} {\bibfnamefont {O.}~\bibnamefont
			{Parcollet}}, \bibinfo {author} {\bibfnamefont {M.}~\bibnamefont {Ferrero}},
		\bibinfo {author} {\bibfnamefont {T.}~\bibnamefont {Ayral}}, \bibinfo
		{author} {\bibfnamefont {H.}~\bibnamefont {Hafermann}}, \bibinfo {author}
		{\bibfnamefont {I.}~\bibnamefont {Krivenko}}, \bibinfo {author}
		{\bibfnamefont {L.}~\bibnamefont {Messio}},\ and\ \bibinfo {author}
		{\bibfnamefont {P.}~\bibnamefont {Seth}},\ }\href
	{http://www.sciencedirect.com/science/article/pii/S0010465515001666}
	{\bibfield  {journal} {\bibinfo  {journal} {Comput. Phys. Commun.}\ }\textbf
		{\bibinfo {volume} {196}},\ \bibinfo {pages} {398} (\bibinfo {year}
		{2015})},\ \bibinfo {note} {version 1.4}\BibitemShut {NoStop}%
	\bibitem [{\citenamefont {Seth}\ \emph {et~al.}(2016)\citenamefont {Seth},
		\citenamefont {Krivenko}, \citenamefont {Ferrero},\ and\ \citenamefont
		{Parcollet}}]{Seth2016}%
	\BibitemOpen
	\bibfield  {author} {\bibinfo {author} {\bibfnamefont {P.}~\bibnamefont
			{Seth}}, \bibinfo {author} {\bibfnamefont {I.}~\bibnamefont {Krivenko}},
		\bibinfo {author} {\bibfnamefont {M.}~\bibnamefont {Ferrero}},\ and\ \bibinfo
		{author} {\bibfnamefont {O.}~\bibnamefont {Parcollet}},\ }\href
	{http://www.sciencedirect.com/science/article/pii/S001046551500404X}
	{\bibfield  {journal} {\bibinfo  {journal} {Comput. Phys. Commun.}\ }\textbf
		{\bibinfo {volume} {200}},\ \bibinfo {pages} {274} (\bibinfo {year}
		{2016})}\BibitemShut {NoStop}%
	\bibitem [{\citenamefont {Bl{\"{o}}chl}\ \emph {et~al.}(1994)\citenamefont
		{Bl{\"{o}}chl}, \citenamefont {Jepsen},\ and\ \citenamefont
		{Andersen}}]{Blochl1994}%
	\BibitemOpen
	\bibfield  {author} {\bibinfo {author} {\bibfnamefont {P.~E.}\ \bibnamefont
			{Bl{\"{o}}chl}}, \bibinfo {author} {\bibfnamefont {O.}~\bibnamefont
			{Jepsen}},\ and\ \bibinfo {author} {\bibfnamefont {O.~K.}\ \bibnamefont
			{Andersen}},\ }\href {http://link.aps.org/doi/10.1103/PhysRevB.49.16223}
	{\bibfield  {journal} {\bibinfo  {journal} {Phys. Rev. B}\ }\textbf {\bibinfo
			{volume} {49}},\ \bibinfo {pages} {16223} (\bibinfo {year}
		{1994})}\BibitemShut {NoStop}%
	\bibitem [{\citenamefont {Krivenko}()}]{TRIQS-ARPACK}%
	\BibitemOpen
	\bibfield  {author} {\bibinfo {author} {\bibfnamefont {I.}~\bibnamefont
			{Krivenko}},\ }\href@noop {} {}\bibinfo {note}
	{\href{https://zenodo.org/record/3930203}{https://zenodo.org/record/3930203}}\BibitemShut
	{NoStop}%
	\bibitem [{\citenamefont {Brinkman}\ and\ \citenamefont
		{Rice}(1970)}]{Brinkman1970}%
	\BibitemOpen
	\bibfield  {author} {\bibinfo {author} {\bibfnamefont {W.~F.}\ \bibnamefont
			{Brinkman}}\ and\ \bibinfo {author} {\bibfnamefont {T.~M.}\ \bibnamefont
			{Rice}},\ }\href {https://doi.org/10.1103/PhysRevB.2.4302} {\bibfield
		{journal} {\bibinfo  {journal} {Phys. Rev. B}\ }\textbf {\bibinfo {volume}
			{2}},\ \bibinfo {pages} {4302} (\bibinfo {year} {1970})}\BibitemShut
	{NoStop}%
	\bibitem [{\citenamefont {Fabrizio}(2007)}]{Fabrizio2007}%
	\BibitemOpen
	\bibfield  {author} {\bibinfo {author} {\bibfnamefont {M.}~\bibnamefont
			{Fabrizio}},\ }\href {https://doi.org/10.1103/PhysRevB.76.165110} {\bibfield
		{journal} {\bibinfo  {journal} {Phys. Rev. B}\ }\textbf {\bibinfo {volume}
			{76}},\ \bibinfo {pages} {165110} (\bibinfo {year} {2007})}\BibitemShut
	{NoStop}%
	\bibitem [{\citenamefont {Slater}(1951)}]{Slater1951}%
	\BibitemOpen
	\bibfield  {author} {\bibinfo {author} {\bibfnamefont {J.~C.}\ \bibnamefont
			{Slater}},\ }\href {https://doi.org/10.1103/PhysRev.82.538} {\bibfield
		{journal} {\bibinfo  {journal} {Phys. Rev.}\ }\textbf {\bibinfo {volume}
			{82}},\ \bibinfo {pages} {538} (\bibinfo {year} {1951})}\BibitemShut
	{NoStop}%
\end{thebibliography}
%apsrev4-2.bst 2019-01-14 (MD) hand-edited version of apsrev4-1.bst
%Control: key (0)
%Control: author (72) initials jnrlst
%Control: editor formatted (1) identically to author
%Control: production of article title (-1) disabled
%Control: page (0) single
%Control: year (1) truncated
%Control: production of eprint (0) enabled
%

\end{document}